\documentclass[10pt,journal,final,twoside]{IEEEtran}

\IEEEoverridecommandlockouts
\usepackage{mysymbol}

\usepackage{xspace,empheq,fancybox,amsmath,amssymb,graphicx,epstopdf,epsfig,syntonly,times,amsthm}
\usepackage{psfrag,color,bm,array,cite}
\usepackage{url,cite,footnote,xspace,syntonly,algorithm,algorithmic}
\usepackage{verbatim,multirow}
\usepackage[T1]{fontenc}
\usepackage{mathtools}
\usepackage{amsmath,amsfonts,amsthm,bm} % Math packages
\usepackage{graphicx}
\usepackage{mwe}
\usepackage{caption}
\usepackage{cite}
\usepackage{amsmath,amssymb,amsfonts}
\usepackage{algorithmic}
\usepackage{graphicx}
\usepackage{textcomp}
\usepackage[caption=false,font=footnotesize]{subfig}
\usepackage{color}
\usepackage{xcolor}
\usepackage{cite}
\usepackage{graphicx}

\usepackage{epstopdf}
\usepackage{comment}
\usepackage{multirow}
\usepackage{paralist}
\usepackage{lineno}
\usepackage{mdwlist}
\usepackage{eurosym}\DeclareGraphicsExtensions{.pdf,.png,.jpg}
\usepackage{breqn}
\usepackage{makecell}
\usepackage{soul}
\usepackage{empheq}
\graphicspath{{./pic/}}
\usepackage[super]{nth}
\usepackage{subfig}
\usepackage{bm}
\usepackage[normalem]{ulem}
\usepackage{xcolor}
\def\BibTeX{{\rm B\kern-.05em{\sc i\kern-.025em b}\kern-.08em
    T\kern-.1667em\lower.7ex\hbox{E}\kern-.125emX}}

\newcommand\upj{\mathord{\mathrm{j}}}
\usepackage{accents}

\newtheorem{remark}{\bfseries Remark}
\newtheorem{proposition}{\bfseries Proposition}

\newcommand{\yc}{\color{teal}{}}
\newcommand{\hao}{\color{magenta}{}}

\allowdisplaybreaks

\usepackage{pifont}% http://ctan.org/pkg/pifont

\theoremstyle{remark}

\DeclarePairedDelimiterX\set[1]\lbrace\rbrace{\def\given{\;\delimsize\vert\;}#1}

\begin{document}
%Include multi-phase / unbalanced?
\title{Distribution Grid Reconfiguration and Restoration via Linearizable Multi-phase Power Flow Modeling}
%\title{Large-Scale Distribution Grid Reconfiguration and Restoration via Linearizable Power Flow Modeling}
%\title{Data-driven Modeling of Linearizable Power Flow for Large-scale Grid Topology Optimization
\author{\IEEEauthorblockN{Young-ho Cho},~\IEEEmembership{Student Member, IEEE}, and \IEEEauthorblockN{Hao Zhu},~\IEEEmembership{Senior Member, IEEE}

\thanks{\protect\rule{0pt}{3mm} This work has been supported by NSF Grants 2130706 and 2150571.} \thanks{\protect\rule{0pt}{3mm} The authors are with the Chandra Family Department of Electrical \& Computer Engineering, The University of Texas at Austin, Austin, TX, 78712, USA; Emails: {\{jacobcho, haozhu\}{@}utexas.edu}.}
}
\maketitle

\begin{abstract}
Accurate and tractable power-flow (PF) modeling is essential for distribution grid topology optimization, where the presence of discrete branch-status decisions therein significantly increase the computational complexity. To tackle this challenge, we develop a topology-adaptive, low-complexity neural piecewise-linear (N-PWL) PF model. We propose a sequential layer design that can directly incorporate branch status variables into the PF mapping from branch flows to nodal injections. As a result, our N-PWL model not only preserves the PF branch-node consistency but also facilitates the representation of any unseen topologies without retraining. A low-complexity model design is further pursued to attain a linear NN parameter scaling by using a fixed-size area partition of the feeder. The resulting scalable N-PWL model enables mixed-integer linear programming (MILP) formulations of feeder topology optimization tasks, including distribution network reconfiguration (DNR) and restoration ordering problems (ROP). Numerical studies on the unbalanced, multi-phase IEEE 123-bus feeder demonstrate our N-PWL model outperforms existing end-to-end baselines in PF prediction and, more importantly, can fill the gap by achieving accurate representation of any unseen topologies. For solving DNR/ROP, our model can significantly improve the AC feasibility over linearized PF models, with a notable reduction of computation time compared with AC-PF based nonlinear formulations, particularly under a large number of topology decisions.
\end{abstract}

\section{Introduction}

% 1. Distribution topology optimization => DNR, ROP
% 2. Opt proxy vs power flow modeling => feasibility purpose
% 3. PF modeling literature review
% 4. Point 1: Topology adaptivity
% 5. Point 2: Low complexity design
% 6. Task: our work

%%%%%%%%%%%%%%%%%5
% 1. Distribution topology optimization => DNR, ROP
\IEEEPARstart{D}{istribution} grid topology flexibility can improve the efficiency and resilience in feeder operations, but solving related optimization problems in real time is very computationally challenging due to the discrete topology decisions.  %introduce substantial computational challenges.
Notable applications include distribution network reconfiguration (DNR)~\cite{baran1989network,bahrami2024dynamic,oh2025sequential} to reduce power losses and other costs, as well as the restoration ordering problem (ROP)~\cite{van2011vehicle,arif2017power,qiu2023power,xu2024adaptive} for post-disturbance service recovery.
For these important applications, determining the discrete branch statuses under grid operating constraints, particularly the AC power-flow (PF) equations, leads to mixed-integer problems that are non-convex and nonlinear.
Moreover, the presence of temporal coupling in multi-period optimization, such as the ROP, further complicates the computation with an increased search space.
Hence, it is imperative to develop tractable approaches to address the nonlinearity of AC-PF when optimizing network configurations for large-scale feeder systems.

% 2. Opt proxy vs power flow modeling => feasibility purpose
Recently, learning-for-optimization using AI advances has been advocated for accelerating power system computation, broadly categorized into two approaches: direct prediction and data-driven PF approximations.
%
%{\hao may need to shorten the first class}
The first approach builds neural networks (NNs) as \textit{optimization proxies} to learn the mapping from optimization inputs to solutions~\cite{pan2020deepopf,liu2023topology}. Although efficient in inference, these proxies face challenges in ensuring solution feasibility and generalizing to \textit{unseen} operating conditions~\cite{chen2023end}. Especially for feeder topology optimization, the combinatorial topology scenarios therein can substantially increase the requirements on training data~\cite{jia2024two}.
%
%{\hao change the next to emphasize the approximation part.}
Instead, the second approach replaces nonlinear PF equations, a major computational bottleneck, with linear~\cite{liu2018data,liu2021data,li2022data,taheri2026optimizing} or piecewise-linear (PWL)~\cite{chehade2026neo,chen2021data,wang2024piecewise} approximations.
 Using these tractable surrogates, grid optimization problems can be solved more efficiently while explicitly retaining the operating constraints.
In particular, ReLU-based NN models with expressive PWL approximations~\cite{kody2022modeling,chevalier2022global} have been preferred for their exact mixed-integer linear reformulations that are solver-friendly.

%{\hao since your table calls DD PF models why not in the paragraphs?}
Despite their improved accuracy, existing data-driven PF approximations still face two major issues for large-scale feeder topology optimization: limited topology adaptivity and growing parameter complexity. 
See the detailed comparisons in Table~\ref{tbl:literature}.
For topology adaptivity, a PF model needs to be capable of representing how branch statuses affect the nodal power balance due to varying branch flows. 
However, existing methods ~\cite{chen2021data,wang2024piecewise,kody2022modeling,chevalier2022global} largely use an \textit{end-to-end} design under a fixed topology scenario, thus ignoring the underlying branch-node coupling. Although some recent work attempts to incorporate the topology information as an additional input~\cite{chen2023topology,duan2026two}, their reliance on an end-to-end structure does not provide any performance guarantees for unseen topology scenarios at inference time.
As for model scalability, the issue of NN parameters growing with the feeder size can critically affect the PF accuracy and generalizability for large feeders.
Although the PWL neural models can generally outperform the simple linear models, their dense layers lead to a \textit{quadratic parameter complexity}.
Together, these limitations highlight the need for new PF models that can adapt to any unseen topology scenarios without an excessive number of NN parameters.

% 6. Task: our work
To this end, we develop a topology-adaptive neural PWL (N-PWL) PF model with low parameter complexity, suitable for distribution grid topology optimization.
One key feature of our N-PWL model is a sequential layer design that allows us to conveniently incorporate branch status variables into the linear mapping from branch flows to nodal injections. This topology-adaptive design not only preserves the PF branch-node consistency, but also allows the N-PWL model to learn the best branch flow models regardless of the topology and thus can generalize to any unseen topologies \textit{without retraining}.
Furthermore, we develop a \textit{low-complexity N-PWL design} by partitioning the feeder into fixed-size areas. Restricting the N-PWL layer connections within each area makes the number of trainable NN parameters grow \textit{linearly} with the feeder size, as opposed to the quadratic growth of fully connected layers. Compared to the latter, the resulting sparse N-PWL model has been shown to improve the accuracy of PF prediction, and more importantly, enhance the PF branch-node consistency significantly over existing end-to-end NN baseline~\cite{kody2022modeling,chevalier2022global}. In addition, the topology-adaptivity of our N-PWL model has been successfully validated by two key applications of feeder topology optimization: distribution network reconfiguration (DNR) and restoration ordering problem (ROP). Its PWL structure allows for mixed-integer linear program (MILP) reformulations of DNR and ROP, demonstrating increased AC feasibility over the linearized PF baseline and a large reduction of computation time compared with AC-PF based nonlinear formulations.
%
%We evaluate PF approximation accuracy against an end-to-end NN baseline~\cite{kody2022modeling,chevalier2022global} using the unbalanced multi-phase IEEE 123-bus feeder~\cite{schneider2017analytic}.
%Importantly, we assess topology adaptivity under unseen network configurations, which the baseline cannot accommodate without retraining.
%The numerical studies further demonstrate improved AC feasibility over conventional linearized PF models and reduced computation time compared with AC-PF-based optimization for DNR and ROP.
In short, our main contributions are three-fold:
\begin{enumerate}
\item We develop a neural PWL (N-PWL) model to attain \textit{accurate, consistent} PF approximations by explicitly preserving the \textit{branch-node coupling} in AC-PF equations.
\item Our N-PWL model achieves \textit{topology adaptivity} by directly incorporating branch status variables into the layer, allowing the trained model to represent \textit{any unseen feeder topology} as validated by DNR/ROP applications.
\item We develop a \textit{low-complexity N-PWL model} using an area-based partition of the feeder to ensure \textit{linear} parameter complexity for \textit{large-scale feeder applications}.
\end{enumerate}

The remainder of the paper is organized as follows.
Section~\ref{sec:formulation} presents the distribution AC-PF model, along with the  DNR and ROP formulations.
Section~\ref{sec:neuralPF} develops the N-PWL PF model, including its training procedure and low-complexity design.
In Section~\ref{sec:MILP_opt}, we discuss the equivalent mixed-integer linear representation of the N-PWL PF model for DNR/ROP reformulations.
Section~\ref{sec:sim} presents the numerical results, and Section~\ref{sec:con} concludes the paper.

\begin{comment}
\begin{table}[t!]
\begin{center}
\caption{Comparisons of data-driven PF models.}
\label{tbl:literature}
\begin{tabular}{c|c|c|c|c}
\hline
Study & \makecell{Parameter\\complexity} & \makecell{Topology\\adaptivity} & \makecell{Optimization\\compatibility} & \makecell{System\\size}\\
\hline
\cite{liu2018data}          & $\ccalO(N)$   & $\times$       & $\checkmark$    & 123 buses \\
\cite{buason2025sample}      & $\ccalO(N)$   & $\times$       & $\checkmark$   & 2,383 buses \\
\cite{kody2022modeling}      & $\ccalO(N^2)$ & $\times$       & $\checkmark$   & 89 buses \\
\cite{chevalier2022global}   & $\ccalO(N^2)$ & $\times$       & $\checkmark$   & 200 buses \\
\cite{jalving2024physics}    & $\ccalO(N^2)$ & $\times$       & $\checkmark$   & 118 buses \\
\cite{cho2023topology}       & $\ccalO(N^2)$ & $\checkmark$   & $\checkmark$   & 118 buses \\
\cite{lin2024powerflownet}   & $\ccalO(N)$   & $\times$       & ? %$\times$       
& 6,470 buses \\
Our work                   & $\ccalO(N)$   & $\checkmark$   & $\checkmark$   & 6,716 buses \\
\hline
\end{tabular}
\end{center}
\end{table}
\end{comment}

\begin{table}[t!]
%\begin{center}
\centering
\caption{Comparisons of data-driven PF models.}% {\hao 19,20 should be ? right?}
\label{tbl:literature}
\begin{tabular}{c|c|c|c|c}
\hline
Study & \makecell{Parameter\\complexity} & \makecell{Topology\\adaptivity} & \makecell{Modeling\\approach} & \makecell{System\\size}\\ 
\hline
\cite{liu2018data} & $\ccalO(N)$ & $\times$ & End-to-end & 123 buses\\
\cite{liu2021data} & $\ccalO(N)$ & $\times$ & End-to-end & 8,500 buses\\
\cite{li2022data} & $\ccalO(N)$ & $\times$ & Sequential & 123 buses\\
\cite{chen2021data} & $\ccalO(N^2)$ & $\times$ & End-to-end & 123 buses\\
\cite{wang2024piecewise} & $\ccalO(N^2)$ & $\times$ & End-to-end & 33 buses\\
\cite{kody2022modeling} & $\ccalO(N^2)$ & $\times$ & End-to-end & 89 buses\\
\cite{chevalier2022global} & $\ccalO(N^2)$ & $\times$ & End-to-end & 200 buses\\
\cite{chen2023topology} & $\ccalO(N^2)$ & \textbf{?} & End-to-end & 123 buses\\
\cite{duan2026two} & $\ccalO(N^2)$ & \textbf{?} & End-to-end & 118 buses\\
Our work & $\ccalO(N)$ & $\checkmark$ & Sequential & 123 buses\\
\hline
\end{tabular}
%\end{center}
\end{table}

\section{Distribution Grid Topology Optimization} \label{sec:formulation}

We present the AC distribution-grid model and specialize it for two topology-optimization tasks: the single-period distribution network reconfiguration (DNR) and multi-period restoration ordering problem (ROP). 
Consider a balanced distribution system with all buses in $\mathcal{N}:=\{1,\ldots,N\}$ and all \textit{potential} branches in $\mathcal{L}:=\{(i,j)\}\subseteq\mathcal{N}\times\mathcal{N}$. The latter can be transformers, voltage regulators, or line segments (closed or open tie lines). For each bus $i\in\mathcal{N}$, let $\mathcal{V}_i:=V_i e^{\upj\theta_i}$ denote its complex voltage phasor, with the magnitude $V_i$ and angle $\theta_i$, respectively.
%Let bus 1 represent the feeder head, providing the reference with $\mathcal{V}_1=1.0 e^{\upj 0}$.
Let bus 1 represent the feeder head and provide the voltage reference $\mathcal{V}_1=V_{\mathrm{ref}}e^{\upj 0}$, where $V_{\mathrm{ref}}$ denotes the reference-voltage magnitude.

\begin{remark}{\rm (Unbalanced multi-phase extension.)}\label{rmk:phase}
Our proposed work is also applicable to unbalanced multi-phase feeders, but for notational simplicity, the majority of the paper presents the balanced single-phase formulation.
%{\hao i switch the order to emphasize on the former}
Specifically, extending \eqref{powerflow} to a multi-phase feeder would replace the scalar voltage and flow variables with phase-indexed vectors, as well as the scalar branch admittances with phase-coupled admittance matrices~\cite{low2025power,bazrafshan2017comprehensive}. Similar to existing linearized models~\cite{gan2014convex,bernstein2018load} developed for unbalanced systems, the proposed model in Section~\ref{sec:neuralPF} supports unequal phase loading, missing phases, and mutual coupling without changing its underlying architecture. The full phase-indexed formulation is omitted for brevity, while Section~\ref{sec:sim} validates this extension using the unbalanced IEEE 123-bus feeder.
\end{remark}

We first introduce a binary status variable $\epsilon_{ij}\in\{0,1\}$ ($0/1$: open/closed) for each line $(i,j)\in\mathcal{L}$, to be used by topology optimization later on. With a switchable $\epsilon_{ij}$, the \textit{effective} complex power $S_{ij}:=P_{ij}+\upj Q_{ij}$ that has been physically transferred from bus $i$ to bus $j$ becomes
\begin{align} \label{actual_flow}
     S_{ij} := \epsilon_{ij} \widehat S_{ij} = \epsilon_{ij}  (\widehat P_{ij}+\upj\widehat Q_{ij})
\end{align}
where $\widehat S_{ij}:=\widehat P_{ij}+\upj\widehat Q_{ij}$ denotes the \textit{potential} line flow. Clearly, an open line with $\epsilon_{ij}=0$ carries zero $S_{ij}$, whereas a closed line follows the AC power flow (PF) equation based on the equivalent  $\pi$-circuit line representation. Specifically, with a given line admittance matrix, applying Ohm's law yields the bi-directional \textit{potential} branch currents as: 
\begin{align*}
\begin{bmatrix}\widehat I_{ij}\\ \widehat I_{ji}\end{bmatrix}
=
\begin{bmatrix}Y_{ij}^{\mathrm{ff}}&Y_{ij}^{\mathrm{ft}}\\
Y_{ij}^{\mathrm{tf}}&Y_{ij}^{\mathrm{tt}}\end{bmatrix}
\begin{bmatrix}\mathcal V_i\\ \mathcal V_j\end{bmatrix}% \label{eq:ohms}
\end{align*}
where the four given admittance values incorporate series and shunt parameters and also tap ratios of transformers or voltage-regulators. While tap ratios are typically fixed in topology optimization problems, it is possible to optimize them together as shown by \cite{robbins2015optimal,bazrafshan2019optimal}. With the similarity between $\widehat S_{ji}$ and $\widehat S_{ij}$, we express below $\widehat S_{ij}$ only:  %{\hao do we need both? i thought it suffices to have just one}
%\begin{subequations}
%\label{powerflow}
\begin{align}
\widehat S_{ij}
&=\mathcal V_i\widehat I_{ij}^{*}
=\left(Y_{ij}^{\mathrm{ff}}\right)^{*}V_i^2
+\left(Y_{ij}^{\mathrm{ft}}\right)^{*}\mathcal V_i\mathcal V_j^{*}.\label{powerflow}
%\widehat S_{ji}
%&=\mathcal V_j\widehat I_{ji}^{*}
%=\left(Y_{ij}^{\mathrm{tt}}\right)^{*}V_j^2
%+\left(Y_{ij}^{\mathrm{tf}}\right)^{*}\mathcal V_j\mathcal V_i^{*}.
%\label{powerflow_2}
\end{align}

Nodal power balance links the effective line flows to nodal distributed generation (DG) and demand. Per bus $i\in\ccalN$, let $S_i^{G}:=P_i^G+\upj Q_i^G$ and $S_i^{D}:=P_i^D+\upj Q_i^D$ denote its local generation and \textit{nominal} demand, respectively. For buses with no DGs, we set $S_i^{G}=0$. Using $\mathcal N_i$ for the subset of buses connected to bus $i$, we write its net power injection as
\begin{align}
S_i &= S_i^{G} - S_i^{D}
= \textstyle\sum_{j\in\mathcal{N}_i} \left(\epsilon_{ij} \widehat S_{ij}\right),
\qquad \forall i\in\mathcal{N}.
\label{powerinj}
\end{align}

\begin{comment}
Nodal power balance links the actual branch flows to distributed generation (DG) and demand. Per bus $i\in\ccalN$, let $\{P_i^{G},Q_i^{G}\}$ and $\{P_i^{D},Q_i^{D}\}$ denote its local generation and nominal demand, respectively. For a bus without any generation, we set $P_i^{G}=Q_i^{G}=0$. Thus, 
the net power injections become
\begin{subequations}
\label{powerinj}
\begin{align}
P_i &= P_i^{G} - P_i^{D}
= \textstyle\sum_{j\in\mathcal{N}_i} P_{ij},
\qquad \forall i\in\mathcal{N},
\label{powerinj_P}\\
Q_i &= Q_i^{G} - Q_i^{D}
= \textstyle \sum_{j\in\mathcal{N}_i} Q_{ij},
\qquad \forall i\in\mathcal{N}.
\label{powerinj_Q}
\end{align}
\end{subequations}
\end{comment}

To facilitate the DNR/ROP formulations, we list the common operating limits here for any energized system, to enforce acceptable voltages, resource capabilities, and line ratings: %{\hao what about V1?}
\begin{subequations}
\label{bound}
\begin{align}
&\underline V_i \leq V_i \leq \bar V_i, \quad V_1=V_{\mathrm{ref}}, \quad
\underline\theta_i \leq\theta_i \leq \bar \theta_i, \quad
\theta_1=0, \label{OPF1}\\
&\underline P_i^{G} \leq P_i^{G} \leq\bar P_i^{G}, \quad
\underline Q_i^{G} \leq Q_i^{G} \leq\bar Q_i^{G}, ~\forall i \in \ccalN \label{OPF2}\\
%&0\leq\widetilde P_i^{D}\leq P_i^{D}, \quad
%0\leq\widetilde Q_i^{D}\leq Q_i^{D}, \label{OPF3}\\
&\underline P_{ij}\epsilon_{ij} \leq P_{ij} \leq\bar P_{ij}\epsilon_{ij}, \quad
\underline Q_{ij}\epsilon_{ij} \leq Q_{ij} \leq\bar Q_{ij}\epsilon_{ij}, \label{OPF3}\\
&P_{ij}^{2}+Q_{ij}^{2}\leq \bar S_{ij}^{2}\epsilon_{ij},~\forall (i,j)\in\ccalL.
\label{OPF4}
\end{align}
\end{subequations}
Constraints~\eqref{OPF1}--\eqref{OPF2} impose bus-wise limits on voltages and controllable resources.
For \eqref{OPF3}--\eqref{OPF4}, they impose branch-wise limits at both terminals, i.e., to both $S_{ij}$ and $S_{ji}$ for the same status $\epsilon_{ij}$.
These constraints will be used to specify the load-service and topology requirements for DNR and ROP.

\subsection{Distribution Network Reconfiguration (DNR)} \label{sec:DNR}
%{\hao it just occured to me when revising intro that DNR not only for loss, so we may want to comment on it after (5a)}
The DNR problem~\cite{baran1989network,bahrami2024dynamic,oh2025sequential} selects a radial feeder configuration to minimize the active power losses while maintaining full load service, by operating controllable sectionalizing and tie switches to transfer loads among alternative feeder paths.
The AC-DNR formulation jointly determines the switch statuses and corresponding feasible operating point. Let $\mathcal{L}^{\mathrm{sw}}\subseteq\mathcal{L}$ denote the subset of controllable lines, and for any $(i,j)\notin \mathcal{L}^{\mathrm{sw}}$, this non-switchable line has its status fixed at $\bar{\epsilon}_{ij}$. Using the actual flows defined in \eqref{actual_flow}, one can form the active power loss of branch $(i,j)$ by $(P_{ij}+P_{ji})$, giving rise to:
\begin{subequations}
\label{DNR}
\begin{align}
    P_{\mathrm{loss}}^*:=\min \quad
    &\textstyle \sum_{(i,j)\in\mathcal{L}}
    \left(P_{ij}+P_{ji}\right)
    \label{DNR_a}\\
    \text{s.t.}\quad
    &\eqref{actual_flow},\,\eqref{powerflow},\,\eqref{powerinj},\,\eqref{bound},
    \label{DNR_b}\\
    &\epsilon_{ij}=\bar{\epsilon}_{ij},
    \qquad
    \forall (i,j)\in\mathcal{L}\setminus\mathcal{L}^{\mathrm{sw}},
    \label{DNR_c}\\
    &\mathcal{G}(\bbepsilon) \text{ is source-connected and radial}.
    \label{DNR_d}
\end{align}
\end{subequations}
The cost in \eqref{DNR_a} focuses on the conventional loss-minimization objective, but it is also possible to add the DG costs and switching penalties; see e.g.,~\cite{zhou2021three,singh2022joint}.
Note that $\mathcal{G}(\bbepsilon)$ represents the topology induced by all closed lines, while \eqref{DNR_c} limits topology decisions to only controllable switches in $\mathcal{L}^{\mathrm{sw}}$. The source-connected and radial conditions in \eqref{DNR_d} are commonly represented by spanning-tree or flow-based constraints~\cite{wang2020radiality,jabr2012minimum}, which are omitted here for brevity. Binary switch decisions and nonlinear AC PF relations make \eqref{DNR} a mixed-integer nonlinear program.

\subsection{Restoration Ordering Problem (ROP)} \label{sec:ROP}
The distribution ROP~\cite{van2011vehicle,arif2017power,qiu2023power,xu2024adaptive} aims to determine the restoration order of damaged lines and the resulting load recovery over a time horizon of \textit{multiple periods}. Following a disruptive event, the ROP maximizes the total energy served over the restoration horizon $\mathcal{T}:=\{1,\ldots,T\}$, as temporary load curtailment is needed under limited repair resources and thus reduced network transferability. 
To formulate the ROP, let $\mathcal{L}^{\mathrm{dam}}\subseteq\mathcal{L}$ denote the given set of damaged lines that were energized prior to the disruptive event, and all other variables including the power demands and line flows are indexed by $t$ for this multi-period problem.
For each $(i,j)\in\mathcal{L}^{\mathrm{dam}}$, we use the decision variable $\epsilon_{ij,t}$ to indicate whether the line has been restored and energized by period $t\in\mathcal{T}$. Thus, $\epsilon_{ij,t}$ jointly represents the repair action and line status under the assumption that a restored line is immediately energized and remains in service. For simplicity, only the statuses of damaged lines are optimized, while the statuses of all remaining lines are fixed at their prescribed values $\bar{\epsilon}_{ij}$.
To further quantify the level of demand recovery in each period, we use  $x_{i,t}\in[0,1]$ to denote the fraction of nominal demand served at bus $i$ in period $t$. 
Under a fixed power factor, we express the \textit{served demand} and update the nodal balance in \eqref{powerinj} per time $ t\in\mathcal{T}$ as
\begin{subequations} \label{ROP_load}
\begin{align}
\widetilde S_{i,t}^{D} &=x_{i,t}S_{i,t}^{D}, \quad
\forall i\in\mathcal{N},\\
S_{i,t} &= S_{i,t}^{G} - \widetilde S_{i,t}^{D} = \textstyle \sum_{j\in\mathcal{N}_i} \left(\epsilon_{ij,t} \widehat S_{ij,t}\right), \quad
\forall i\in\mathcal{N}.
\end{align}
\end{subequations}

Given the post-event initial $\epsilon_{ij,0}$ and the budget $\eta$ on the maximum number of damaged lines restored per period, we formulate the AC-ROP as:
\begin{subequations}
\label{ROP}
\begin{align}
E_{\mathrm{srv}}^*:=\max \quad
&\textstyle \sum_{t\in\mathcal{T}}\textstyle \sum_{i\in\mathcal{N}} \left(x_{i,t}P_{i,t}^{D}\right)
\label{ROP_a}\\
\text{s.t.}\quad
&\eqref{actual_flow},\,\eqref{powerflow},\,\eqref{bound},\,\eqref{ROP_load},
\label{ROP_b}\\
%&S_{i,t} = S_{i,t}^{G} - \widetilde S_{i,t}^{D} = \textstyle \sum_{j\in\mathcal{N}_i} S_{ij,t},
%\quad \forall i\in\mathcal N,
%\label{ROP_c}\\
&\epsilon_{ij,t}=\bar{\epsilon}_{ij},
\quad \forall (i,j)\in \mathcal{L}\setminus\mathcal{L}^{\mathrm{dam}}, \label{ROP_e}\\
&\epsilon_{ij,t-1}\leq\epsilon_{ij,t},
\quad \forall (i,j)\in\mathcal{L}^{\mathrm{dam}}, \label{ROP_f}\\
&\textstyle \sum_{(i,j)\in\mathcal{L}^{\mathrm{dam}}} \left(\epsilon_{ij,t}-\epsilon_{ij,t-1}\right) \leq\eta, \label{ROP_g}\\
&\epsilon_{ij,T}=1,
\quad \forall (i,j)\in\mathcal{L}^{\mathrm{dam}}, \label{ROP_h}\\
&x_{i,t-1}\leq x_{i,t}, \quad \forall i\in\mathcal{N},\ t\in\mathcal{T}, \label{ROP_i}\\
&\mathcal{G}(\bbepsilon_t) \text{ is source-connected and radial}.
\label{ROP_j}
\end{align}
\end{subequations}
Constraints~\eqref{ROP_e}--\eqref{ROP_h} together fix the statuses of undamaged lines, maintain restored lines in service, impose the per-period restoration budget, and require a complete restoration within $\ccalT$. Moreover, constraints~\eqref{ROP_i}--\eqref{ROP_j} enforce a monotonic load recovery, and the source-connected radial operation similar to \eqref{DNR_d}.
Clearly, the time-coupled nature of \textit{multi-period} AC-ROP further increases the complexity over the DNR problem. Nevertheless, the varying topology in both problems makes it necessary to keep the AC-PF model, and the resulting nonlinearity can greatly challenge solving these mixed-integer problems. The following section will address this computational burden by developing tractable PF approximations.

\begin{comment}
\label{ROP}
\begin{align}
    \max \quad 
    &\sum_{t\in\mathcal{T}} \sum_{i\in\mathcal{N}} x_{i,t}P_{i,t}^{D}
    \label{ROP_a}\\
    \mathrm{s.t.}\quad
    &\eqref{actual_flow},\,\eqref{powerflow},\,\eqref{powerinj},\,\eqref{bound},\,\eqref{ROP_load},
    \label{ROP_b}\\
    &\sum_{(i,j)\in\mathcal{L}} \epsilon_{ij,t} \leq R_t, \label{ROP_c}\\
    &R_t=R_{t-1}+\eta,~\epsilon_{ij,t-1}\leq\epsilon_{ij,t},~\epsilon_{ij,T}=1,\label{ROP_d}\\
    &\mathcal{G}(\bbepsilon_t) \text{ is connected and radial}. \label{ROP_e}
\end{align}
\end{comment}

\section{Piecewise Linear (PWL) PF Modeling via Neural Approximation} \label{sec:neuralPF}
% Overview
To facilitate feeder topology optimization, we develop a neural PWL (N-PWL) PF model that can attain high approximation accuracy under varying grid topologies. The key innovation in our neural design lies in preserving the PF branch-node \emph{consistency} in \eqref{actual_flow}--\eqref{powerinj}. First, our design decouples the topology dependence from trainable model parameters, making it possible to incorporate any \textit{unseen topology changes} without retraining. Second,  it also matches the nodal balance between branch flows and nodal injections, improving the accuracy of PF prediction. This PF-consistent architecture differs from conventional end-to-end approaches~\cite{kody2022modeling,chevalier2022global} that separately predict flows and injections, ignoring their underlying coupling. We first present the N-PWL architecture and its topology adaptivity, followed by the training procedure and a low-complexity extension for large-scale feeders.

\begin{figure}[t!]
    %\begin{center}
    \includegraphics[scale=0.28]{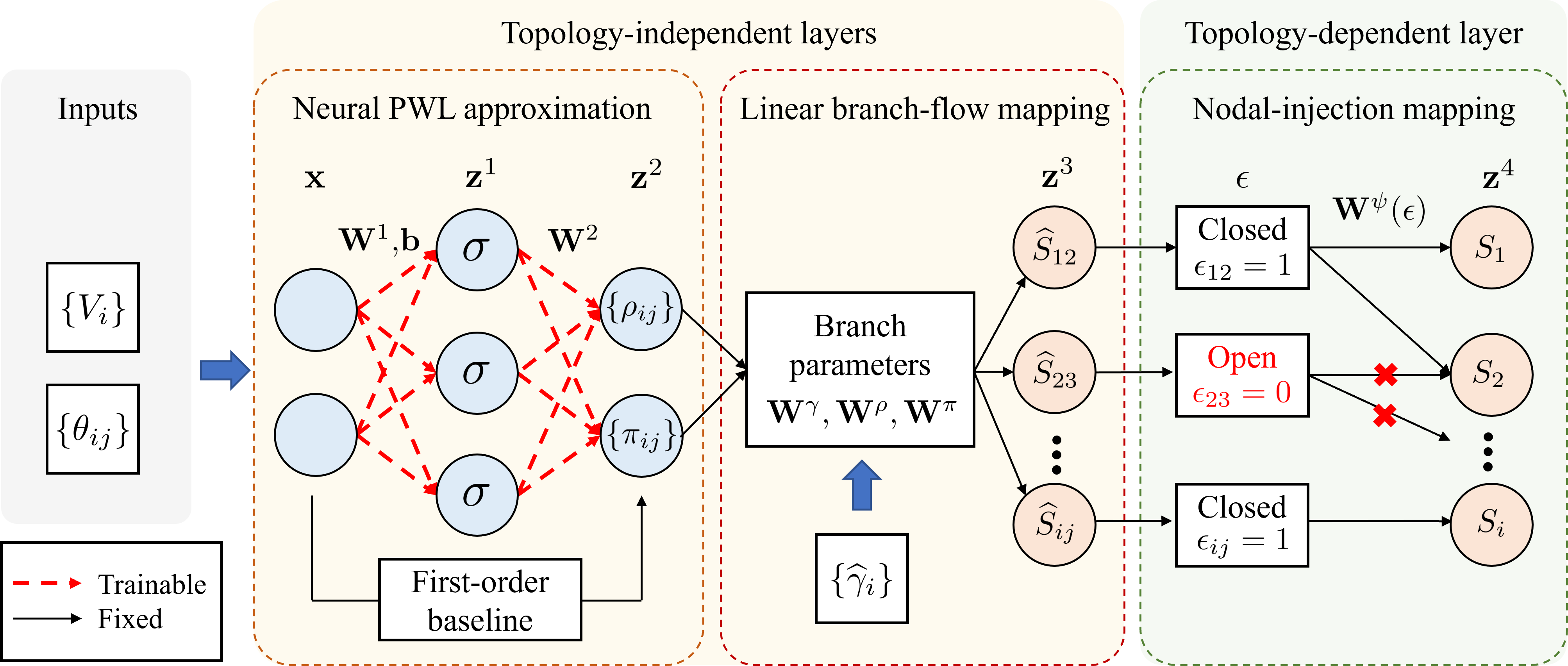}
    \caption{\small The sequential layer design of our proposed N-PWL model: from input voltage states to voltage-product variables via trainable layers, followed by linear fixed layers to PF variables including the topology-dependent final layer.}%{\hao cant tell the dashed line or dotted line}
    \label{structure}
    %\end{center}
\end{figure}

We design the sequential layers for our N-PWL model, shown in Fig.~\ref{structure}, to match the AC-PF model in \eqref{actual_flow}-\eqref{powerinj}. %{\hao fig need to update a few places}
Using the voltage states or PF input in $\bbx$ , the first two layers of  $\{\bbz^1,\bbz^2\}$ seek to approximate the \textit{quadratic, voltage-product terms} denoted by $\bbf(\bbx):=[\bbrho;\bbpi]$, to be detailed soon.
This step also incorporates a first-order baseline specified at a given voltage reference $\check{\bbx}$ with its PF Jacobian matrix $\check{\bbJ}$.
The third layer constructs all potential branch flows in  $\bbz^3$, based on an additional approximation $\widehat \bbgamma\approxeq \bbV^2$.
The final layer forms all nodal injections in $\bbz^4$ according to the line status vector $\bbepsilon$.
Note that model training is not needed for the last two layers, since $\{\bbW^\gamma,\bbW^\rho,\bbW^\pi\}$ are fixed by branch parameters as in \eqref{powerflow} while $\bbW^\psi(\bbepsilon)$ are fully determined by $\bbepsilon$ per \eqref{powerinj}.
%{\hao just realize the new balance (3) doesn't have epsilon, shall we add epsilon in to (3)?}
Thus, it suffices to learn the remaining parameters $\{\bbW^1,\bbW^2,\bbb\}$ during training. Using an activation function $\sigma(\cdot)$, the overall neural model becomes:
\begin{subequations}\label{NN_pf}
\begin{align}
    \bbz^{1}
    &=\sigma\!\left(\bbW^{1}(\bbx-\check{\bbx})+\bbb\right),
    \label{NN_pf1}\\
    \bbz^{2}
    &=\bbW^{2}\bbz^{1}
    +\left(\bbf(\check{\bbx})+\check{\bbJ}(\bbx-\check{\bbx})\right),
    \label{NN_pf2}\\
    \bbz^{3}
    &=\bbW^{\gamma}\widehat \bbgamma
    +[\bbW^{\rho}~\bbW^{\pi}]\bbz^{2},
    \label{NN_pf3}\\
    \bbz^{4}
    &=\bbW^{\psi}(\bbepsilon)\bbz^{3}.
    \label{NN_pf4}
\end{align}
\end{subequations}

For PF consistency, our model leverages a sequential layer design that first predicts voltage-product terms $[\bbrho;\bbpi]$, instead of directly forming PF outputs. This is inspired by the \textit{linear mapping} in \eqref{powerflow}, from the following variables to branch flows:
\begin{subequations}\label{voltage_product}
\begin{align}
\gamma_i&:=V_i^2,\label{self_product}\\
\rho_{ij}+\upj\pi_{ij}
&:=\mathcal V_i\mathcal V_j^{*} = V_iV_j\cos\theta_{ij}+\upj V_iV_j\sin\theta_{ij},\label{cross_product}
\end{align}
\end{subequations}
where $\theta_{ij}:=\theta_i-\theta_j$ is included by the input $\bbx$ in Fig.~\ref{structure}. Note that these quadratic terms are sufficient for constructing all PF outputs in $\{\bbz^3,\bbz^4\}$, as the last two layers are fixed by system parameters. This way, our sequential layers do not depend on line statuses and thus can adapt to any unseen topology, as formally stated in the following proposition.

\begin{proposition}{\rm (Topology adaptivity.)}
\label{prop:topology}
The N-PWL PF model in \eqref{NN_pf}, once trained, can update  $\bbW^\psi(\bbepsilon)$ to directly incorporate any feasible topology $\bbepsilon=[\{\epsilon_{ij}\}]$, even those \textbf{unseen} by the training stage. 
\end{proposition}

Proposition \ref{prop:topology} is valid since the N-PWL model  \eqref{NN_pf} can form $\bbz^3$ to approximate all potential branch flows, regardless of the actual topology or $\bbepsilon$. The same principle allows one to also incorporate any possible changes of branch parameters by updating $\{\bbW^\gamma,\bbW^\rho,\bbW^\pi\}$ in \eqref{NN_pf3}. More importantly, Proposition  \ref{prop:topology} states that the topology update does not need any re-training, but we do require the training dataset to cover the full operating range for all candidate topologies. To achieve this, our training stage will select \textit{representative topologies} such that each branch is energized at least once rather than enumerating all feasible configurations, as discussed in Section~\ref{sec:sim}.

Another design choice is embedding a first-order baseline approximation to improve the overall accuracy. This baseline provides a high-quality local approximation at a given  $\check{\bbx}$, facilitating the prediction of the residual term as advocated by the well-known ResNet framework~\cite{he2016deep,simpson2017theory}.
Accordingly, our N-PWL model follows the neural PF literature \cite{kody2022modeling,chevalier2022global,cho2023topology} to embed the linear PF approximation of $(\bbf(\check{\bbx})+\check{\bbJ}(\bbx-\check{\bbx}))$ in \eqref{NN_pf2}, while the neural correction by $\bbW^2\bbz^1$  targets large, nonlinear residuals due to shifting operating points or topology changes. %can shift the operating point from $\check{\bbx}$ and, especially, alter the angle differences $\theta_{ij}$.
%For $\{\bbrho,\bbpi\}$, these angle shifts can increase the baseline error, so 
Similarly for the layer \eqref{NN_pf3}, we also use a simple first-order approximation  to form  $\widehat\bbgamma$, which is very accurate in practice due to the voltage regulation~\cite{trodden2013optimization}, given by 
\begin{align*}
\widehat\bbgamma
&=2\operatorname{diag}(\check{\bbV})\bbV-\check{\bbgamma},
\end{align*}
with $\check{\bbgamma}:=\check{\bbV}\odot\check{\bbV}$ and $\odot$ denotes element-wise multiplication. When $\check{\bbV}=\mathbf 1$, it reduces to the standard $\widehat\bbgamma=2\bbV-\mathbf 1$.

Last, the activation neuron $\sigma(\cdot)$ in Fig.~\ref{structure} determines the functional form of the N-PWL model. Without any nonlinear activation, the model boils down to a first-order approximation with trainable parameters. Using the ReLU activation ensures a PWL approximation of AC-PF that facilitates the MILP reformulation in Section~\ref{sec:MILP_opt}. The role and scaling of ReLU neurons are discussed in the ensuing training procedure.

\subsection{Training of the N-PWL PF Model} \label{sec:piecewise}
The N-PWL PF model is trained with all available outputs to achieve high accuracy across all layers. Let $\bbz^{\rm pf}$ and $\bbz^{\rm inj}$ denote the target labels of potential branch-flow and nodal-injection vectors, respectively. %; and similarly, use $\bbf(\bbx)$ to represent the actual values for $[\bbrho;\bbpi]$ in \eqref{cross_product}.
The loss function for all trainable parameters $\{\bbW^1,\bbW^2,\bbb\}$ is given by
\begin{align}
    \mathcal{J}(\bbW^1,\bbW^2,\bbb)
    ={}&
    \left\| \bbf(\bbx)-\bbz^2\right\|_2^2
    \nonumber\\
    &+\lambda
    \left\|
    [\bbz^{\mathrm{pf}};\bbz^{\mathrm{inj}}]
    -[\bbz^3;\bbz^4]
    \right\|_2^2,
    \label{update}
\end{align}
which is averaged over all training samples. Note that the hyperparameter $\lambda>0$ seeks to balance the two error terms in \eqref{update}, with a larger value placing higher emphasis on the PF outputs. Selecting $\lambda$ can use standard procedures of cross-validation and hyperparameter tuning.  

\begin{figure}[t!]
    \vspace{-1.5em}
	\centering
	\subfloat[]{\includegraphics[scale=0.29]{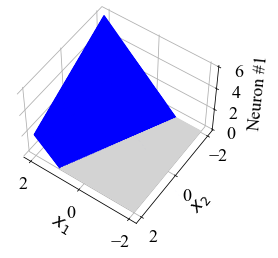}\label{ReLU_a}}
    \subfloat[]{\includegraphics[scale=0.29]{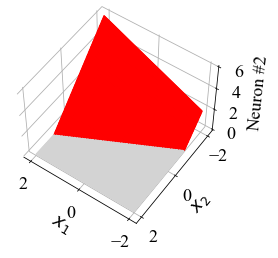}\label{ReLU_b}}
	\subfloat[]{\includegraphics[scale=0.29]{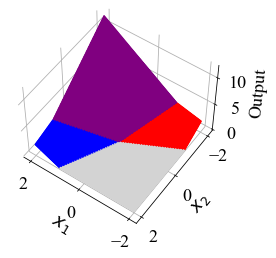}\label{ReLU_c}}
	\caption[]{\small Each ReLU neuron in (a)--(b) produces two affine regions. Their combined output in (c) forms four regions corresponding to the neurons' activation patterns.}\label{ReLU_fig}
\end{figure}

We further discuss the role of ReLU activation in forming the PWL-PF model. The activation status of each hidden neuron in \eqref{NN_pf1} essentially divides the input space into two affine regions. Together, multiple neurons combine individual activation patterns into a complex PWL mapping.
Fig.~\ref{ReLU_fig} illustrates this combination process using two neurons for a two-dimensional input, which together produce up to four activation patterns.
In general, a neural layer with $K$ ReLU neurons admits at most $2^K$ activation patterns (see Remark~\ref{rmk:numberK} for the selection of $K$). This PWL mapping extends to the PF outputs, as formally stated in the following proposition.

\begin{proposition}{(\rm PWL mapping.)}
\label{prop:pwl}
For a fixed topology $\bbepsilon$, using the ReLU activation  for \eqref{NN_pf1} and an affine approximation of $\widehat \bbgamma$ ensures all outputs  $\{\bbz^2,\bbz^3,\bbz^4\}$ in \eqref{NN_pf} are continuous piecewise-linear (PWL) functions of the voltage state $\bbx$.
\end{proposition}

Proposition~\ref{prop:pwl} holds for \eqref{NN_pf} because of its layer-wise composition of affine and ReLU-based mappings. ReLU activation in \eqref{NN_pf1} makes $\bbz^1$ a continuous PWL function of $\bbx$, and by deduction, all subsequent layers remain continuous and PWL due to them being affine mappings.

\begin{figure}[t!]
	\centering
    %\begin{center}
	\includegraphics[width=0.8\linewidth]{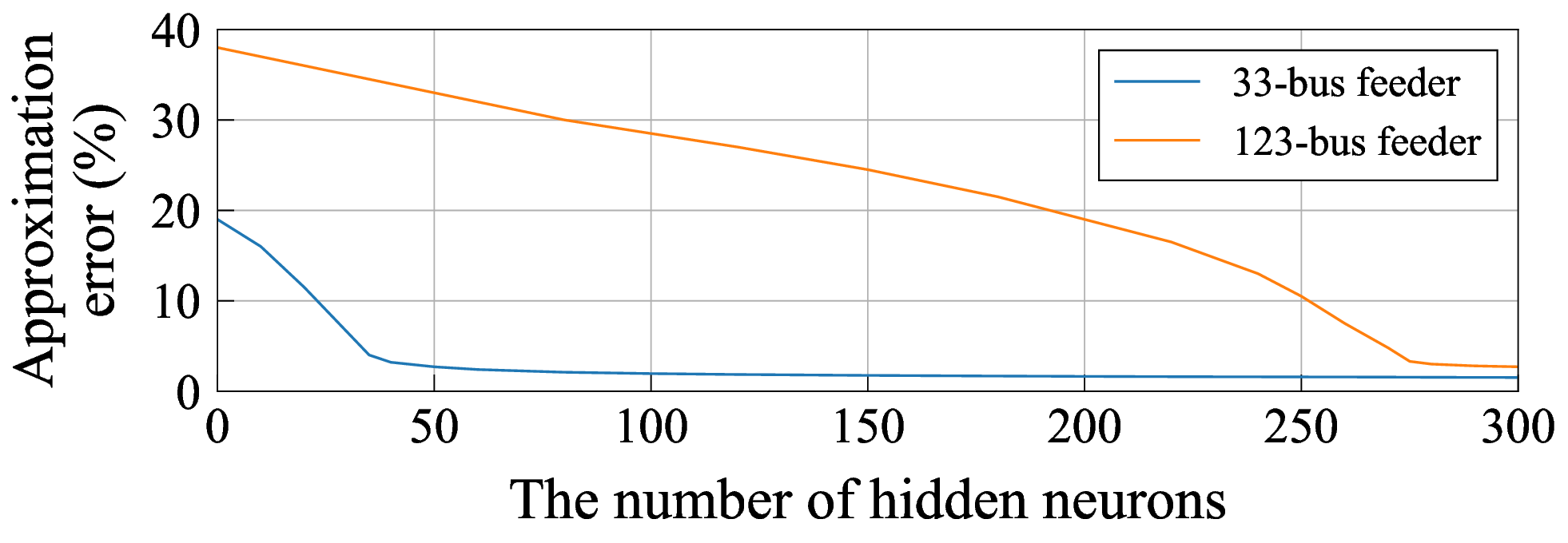}
	\caption{\small Error performance versus the number of hidden neurons for the single-phase IEEE 33-bus and multi-phase IEEE 123-bus feeder systems.} \label{SelectionQ}
	%\end{center}
\end{figure}

\begin{remark}{(\rm Selecting the number of neurons $K$.)} 
\label{rmk:numberK}
A large $K$ enhances the expressiveness of our N-PWL model, but also increases the number of trainable parameters and may affect the model generalizability under limited training data~\cite{trunk1979problem}. Note that the dimensions of the N-PWL input/output scale with the total number of buses, or more precisely, the number of phase nodes, $N_{\phi}$. This makes the scaling of $K$ to be potentially proportional to $N_{\phi}$. 
To examine this scaling, we evaluate the N-PWL PF model using two feeders, the single-phase IEEE 33-bus~\cite{baran1989network} and multi-phase IEEE 123-bus~\cite{schneider2017analytic} systems, with $N_{\phi}=33$ and $274$, respectively.
Fig.~\ref{SelectionQ} plots the average prediction error versus $K$, showing that $K=35$ is adequate for the 33-bus feeder while $K=275$ for the 123-bus feeder.
These results suggest that $K\approxeq N_{\phi}$ could be a proper choice.
\end{remark}

Using $K\approxeq N_{\phi}$, we can show that the parameters of our N-PWL model have a quadratic growth with $N_\phi$.
For a radial feeder, the total number of phase-specific branches $L_{\phi}$ scales linearly with $N_{\phi}$. Using both $L_{\phi}=\ccalO(N_{\phi})$ and $K=\ccalO(N_{\phi})$ from Remark~\ref{rmk:numberK}, the trainable weight matrices $\{\bbW^1,\bbW^2\}$ have $\ccalO((N_{\phi}+L_{\phi})K)=\ccalO(N_{\phi}^2)$ parameters.
This quadratic complexity could challenge large-scale feeder modeling, motivating a scalable model design next.

\subsection{Low-Complexity Modeling for Scalability}\label{sec:sparse}
\begin{figure}[t!]
	\centering
	\subfloat[]{\includegraphics[scale=0.4]{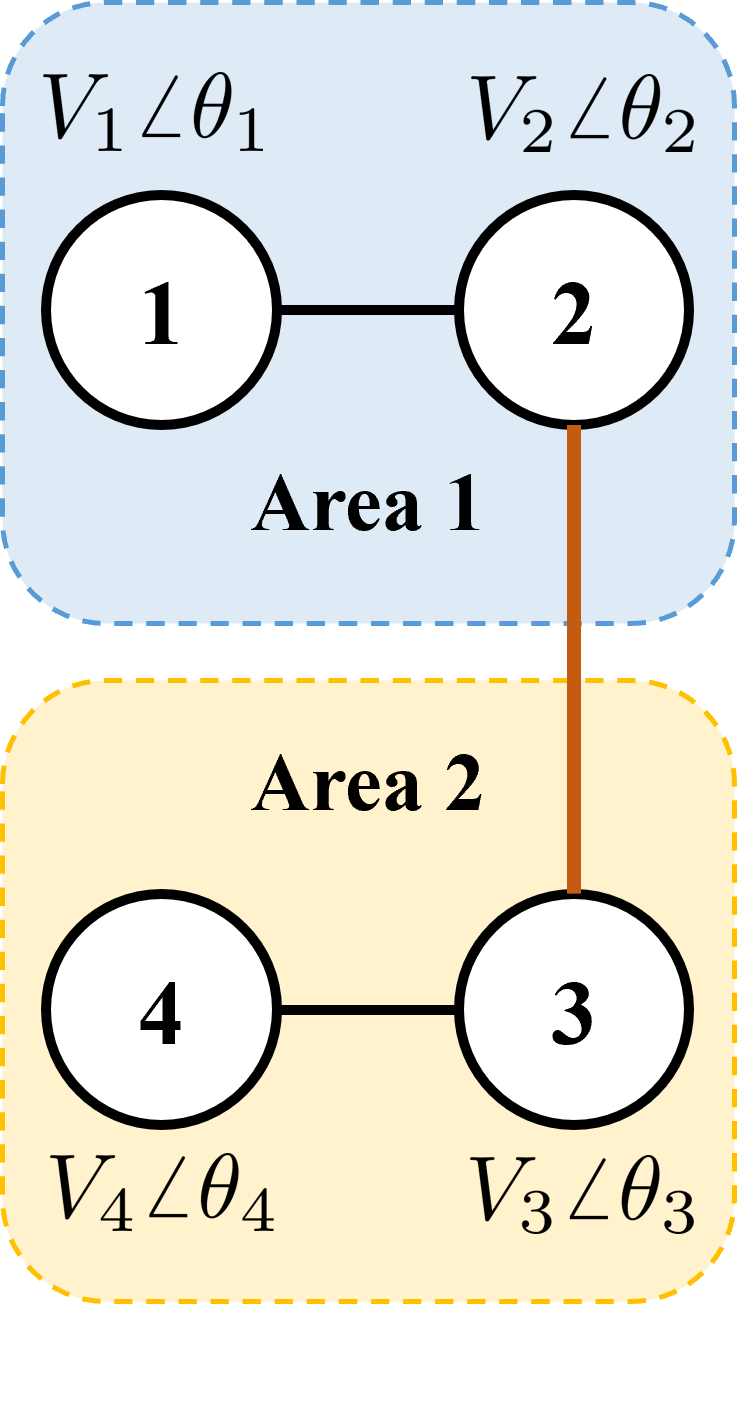}}
	\qquad
	\subfloat[]{\includegraphics[scale=0.4]{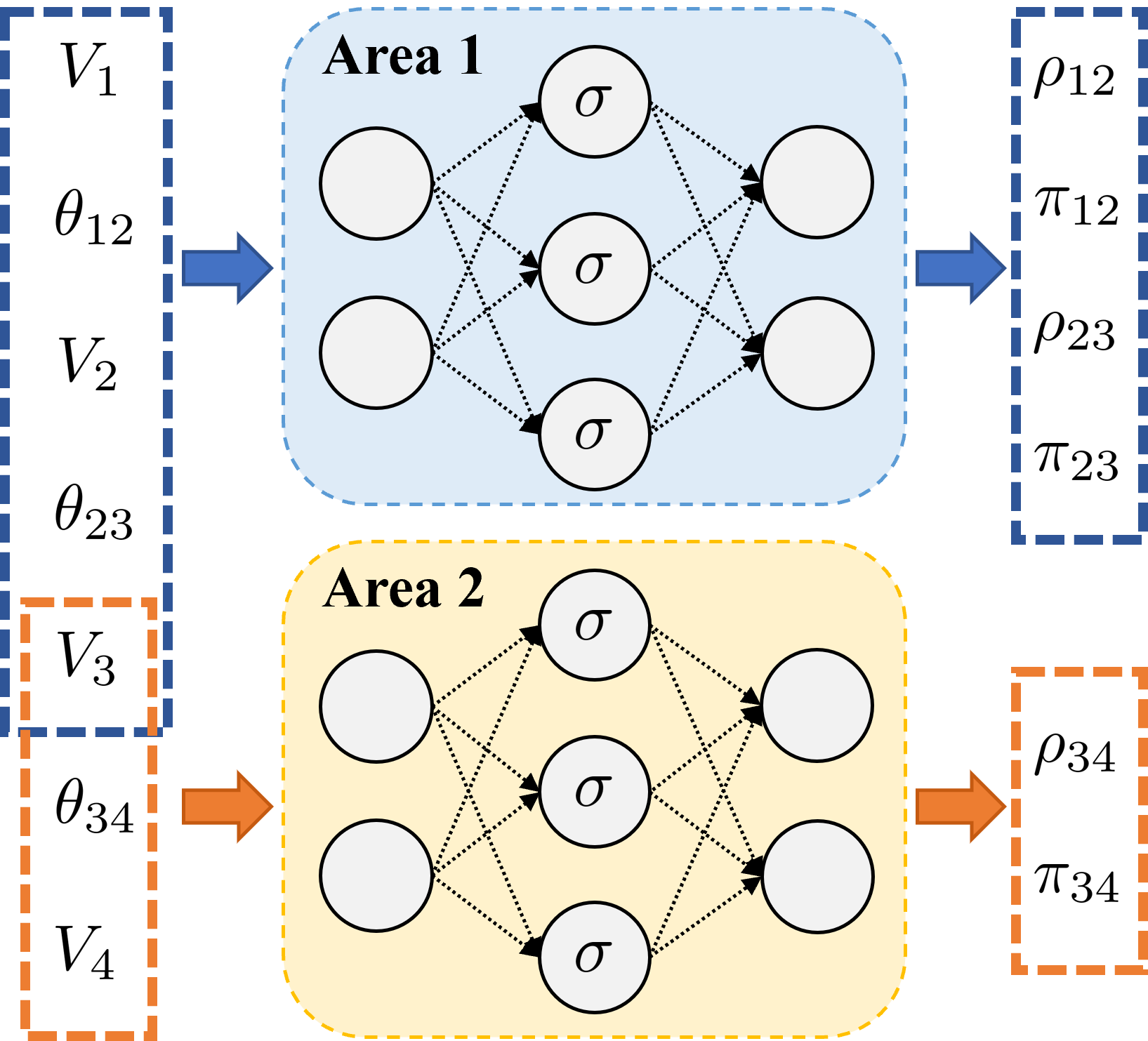}}
	\caption[]{\small (a) A 4-bus feeder partitioned into two areas with one boundary branch, and (b) the corresponding sparse neural architecture. The boundary branch $(2,3)$ is assigned to Area 1 by augmenting its input with the voltage variables at bus~3.}
    \label{sparseGenNN}
\end{figure}

We propose to reduce the model complexity by sparsifying the first two trainable layers \eqref{NN_pf1}--\eqref{NN_pf2} using an area-based design. Note that the layer \eqref{NN_pf4} is unchanged, ensuring the resultant sparse N-PWL to remain topology adaptive. Our approach exploits the \textit{locality} of the voltage-product terms, as each pair $\{\rho_{ij},\pi_{ij}\}$ depends only on the voltages of buses $i$ and $j$.  By partitioning the feeder into multiple areas, we restrict the neural mapping to be within each area only. Fig.~\ref{sparseGenNN} illustrates this idea using a 4-bus feeder, where Area 1 also includes the boundary branch $(2,3)$ and its voltage states $V_3$ and $\theta_{23}$. For each area, its own PF outputs are predicted from a subset of voltage states only.
Overall, this area-based design reduces trainable parameters while retaining the same properties as the fully-connected N-PWL.

To obtain an area-based N-PWL model, we group buses that are \textit{electrically close} into the same area.
This way, each area tends to share similar voltage variability~\cite{luan2024improved}, allowing one mapping to capture correlated variations in its voltage-product terms.
We introduce the metric of \textit{closeness} using the branch affinity $\lambda_{ij}=1/|z_{ij}|$, with the series impedance $z_{ij}$ under the balanced representation~\cite{sanchez2014hierarchical}. This allows us to adopt spectral clustering~\cite{ng2001spectral} to obtain a partition with the intra-area affinities much larger than inter-area ones. In addition, we select the number of areas $C$ close to $N_{\phi}/N'_{\phi}$, such that the number of phase nodes in each area $N_\phi'\ll N_\phi$. This choice allows us to bound the overall model complexity below.

%zhu2012sparse

\begin{proposition}{\rm (Linear model complexity.)}
\label{prop:sparse}
By allocating $N_\phi'\ll N_\phi$ nodes into each area, the sparse N-PWL model has a linear complexity scaling with $N_\phi$, with $\ccalO(N_\phi N'_\phi)$ trainable parameters.
\end{proposition}

Proposition~\ref{prop:sparse} follows from the earlier analysis that each area with $N_\phi'$  phase nodes would have $\ccalO((N'_\phi)^2)$ trainable parameters, giving rise to a total of $\ccalO(C (N'_\phi)^2) = \ccalO(N_\phi N'_\phi)$ ones.  This linear scaling in $N_\phi$ reduces from the previous quadratic order, and thus it can enhance the scalability for large feeder systems.
%The trainable parameters in each area scale as $\ccalO((N'_\phi)^2)$. Since $C$ scales as $N_\phi/N'_\phi$, the total parameter count becomes $\ccalO(N_\phi N'_\phi)$.}

\begin{comment}
{\hao why rmk? how diff from prop 3???}
\begin{remark}{(\rm Parameter reduction by area partitioning.)}
\label{rmk:param_reduction}
{\yc For an approximately balanced $C$-area partition, each area contains about $1/C$ of the system-wide variables and hidden neurons. Thus, each area-specific mapping has roughly $1/C^2$ of the fully connected parameters, yielding approximately $1/C$ of the total parameters across all $C$ areas.
For the IEEE 123-bus feeder, we use five areas with an average of 54.8 phase nodes per area and distribute the total $K=275$ hidden neurons as 55 neurons per area.
This five-area configuration gives an approximate five-fold parameter reduction relative to the fully connected model.}
\end{remark}
\end{comment}

\section{N-PWL Model for Topology Optimization} \label{sec:MILP_opt}
To use our N-PWL PF model for topology optimization, we express it into a mixed-integer linear (MIL) form. This will be used for the MIL program (MILP) reformulations of DNR and ROP, which are efficiently solvable by existing optimizers.

The MIL form follows directly from the use of ReLU neurons in our N-PWL model. Note that the sole source of nonlinearity within  \eqref{NN_pf1}--\eqref{NN_pf3} lies in the ReLU activation in \eqref{NN_pf1}; all subsequent mappings are affine upon fixing $\bbz^1$. Thus, we rewrite \eqref{NN_pf1} using an auxiliary binary vector $\bbbeta$ as the ReLU activation indicator, given by
\begin{align}
    \bbz^1 = \sigma(\bbW^1 \tdbbx + \bbb) =  (\bbW^1 \tdbbx + \bbb) \odot \bbbeta, \label{eq:prod}
\end{align}
where $\tdbbx:=\bbx-\check{\bbx}$.
Instead of the conventional big-$M$ relaxation \cite{griva2009linear} of \eqref{eq:prod}, we apply a tight formulation developed in~\cite{grimstad2019relu} to obtain its MIL form. The latter consists of two major steps: i) using tight pre-activation bounds from the input range; and ii) eliminating neurons whose activation statuses are fixed in that range.

First, tight pre-activation bounds can improve the computational efficiency of the resulting MILP~\cite{fattahi2018bound}. To this end, we first construct the input range for  $\tdbbx\in[\underline{\bbx},\bar{\bbx}]$ based on system operating limits. Using this range, one can possibly bound $(\bbW^1\tdbbx+\bbb)$ by formulating another optimization problem as in~\cite{grimstad2019relu}, but this incurs additional computation. Instead, we adopt the direct sign-based approach in~\cite{anderson2020strong}. Let $n_x:=\dim(\bbx)$ and form two matrices $\{\bbL,\bbU\}\in\mathbb{R}^{n_x\times K}$ according to the entries in $\bbW^1$:
\begin{align}\label{input_bounds}
    \{L_{nk},U_{nk}\} =\begin{cases} \{\underline {x}_n, \bar {x}_n \} & \text{if $ W^1_{kn} \geq 0$}\\
    \{\bar {x}_n, \underline {x}_n\} & \text{if $ W^1_{kn} < 0$} \end{cases}.
\end{align}
Using these two matrices, we can express the $k$-th entry in \eqref{eq:prod}, namely $z_k^1$, via four linear inequalities: 
\begin{subequations}\label{MILP}
\begin{align}
    0 &\leq z^1_k \leq (\bbW^1_{k:} \bbU_{:k}\! +\! b_k) \beta_k,\label{beta0} \\
    \bbW^1_{k:}\tdbbx\!+\!b_k &\leq z^1_k \leq \bbW^1_{k:} (\tdbbx\! -\! \bbL_{:k}(1\!-\!\beta_k))\!+\! b_k \beta_k,\label{beta1}
\end{align}
\end{subequations}
where $\bbW^1_{k:}$ denotes the $k$-th row of $\bbW^1$. If $\bbW^1_{k:}\tdbbx+b_k>0$, constraints~\eqref{beta0}--\eqref{beta1} together enforce $\beta_k=1$ and $z_k^1=\bbW^1_{k:}\tdbbx+b_k$. Otherwise, we have $\beta_k=0$ and $z_k^1=0$. Thus, the four inequalities in \eqref{beta0}--\eqref{beta1} are equivalent to \eqref{eq:prod} when the input range is given. As shown in~\cite{anderson2020strong}, these sign-dependent bounds provide a tighter formulation than the conventional big-$M$ relaxation.

Second, the range $[\underline{\bbx},\bar{\bbx}]$ can directly determine certain $\beta_k$ values and thus eliminate the corresponding inequalities in \eqref{MILP}, further improving computation efficiency. For each $z_k^1$, if the ReLU activation status remains unchanged over the entire range, then we can simply fix $\beta_k$ to 0 or 1. Specifically, if $\bbW^1_{k:}\bbU_{:k}+b_k<0$ holds, then we have $\beta_k=0$ and $z_k^1=0$ for all $\tdbbx$, rendering the inequalities in \eqref{beta1} inactive. Similarly, if $\bbW^1_{k:}\bbL_{:k}+b_k\geq0$, then the resultant equalities of $\beta_k=1$ and $z_k^1=\bbW^1_{k:}\tdbbx+b_k$ make \eqref{beta0} inactive. Hence, we define two subsets of neurons for $\mathcal{K}:=\{1,\ldots,K\}$:
\begin{subequations}\label{ReLUPrune}
\begin{align}
    \mathcal {K}_0 &= \set[\Big] {k \in \mathcal K \given \bbW^1_{k:}\bbU_{:k} + b_k < 0},\\
    \mathcal {K}_1 &= \set[\Big] {k \in \mathcal K \given \bbW^1_{k:}\bbL_{:k} + b_k \geq 0}.
\end{align}
\end{subequations}
For any neuron $k\in\mathcal{K}_0$, it suffices to fix its $\beta_k=0$ and use only \eqref{beta0}; similarly, for any $k\in\mathcal{K}_1$, we have $\beta_k=1$ and \eqref{beta1} only. This preprocessing step of eliminating inactive inequalities does not affect the \textit{optimality }at all, but has been shown to achieve a reduction of approximately $5\%$ of the neurons by our numerical experiments in Section~\ref{sec:sim}.

\begin{proposition}{(\rm MIL representation of N-PWL PF)}
\label{prop:MILP}
With the bounds $\tdbbx\in[\underline{\bbx},\bar{\bbx}]$, the N-PWL potential-flow mappings, once trained, admit the following \textit{equivalent} MIL representation:
\begin{subequations}\label{PWLPF}
\begin{align}
    \bbW^1_{k:}&\tdbbx\!+\!b_k \leq z^1_k \leq \bbW^1_{k:} (\tdbbx\! -\! \bbL_{:k}(1\!-\!\beta_k))\!+\! b_k \beta_k,\nonumber\\
    &\qquad \qquad \qquad \qquad\qquad~~~~\forall k \in \mathcal K - \mathcal K_0\label{PWLPF1}\\
    0 &\leq z^1_k \leq (\bbW^1_{k:} \bbU_{:k}\! +\! b_k) \beta_k,~\forall k \in \mathcal K - \mathcal K_1\label{PWLPF2}\\
    \beta_k &= \begin{cases} 1,~  k \in \mathcal{K}_1\\
    0,~ k \in \mathcal{K}_0 \end{cases}\label{PWLPF3}\\
    [ \widehat\bbrho; \widehat\bbpi]&=[\check{\bbrho};\check{\bbpi}]+ \check {\bbJ} \tdbbx + \bbW^2 \bbz^{1}\label{PWLPF4}\\
    \widehat\bbgamma &= 2\operatorname{diag}(\check{\bbV})\bbV-\check{\bbgamma}\label{PWLPF5}\\
    \widehat S_{ij}
    &=\left(Y_{ij}^{\mathrm{ff}}\right)^{*}\widehat\gamma_i
    +\left(Y_{ij}^{\mathrm{ft}}\right)^{*}
    \left(\widehat\rho_{ij}+\upj\widehat\pi_{ij}\right),
    ~\forall (i,j). \label{PWLPF6}
\end{align}
\end{subequations}
\end{proposition}

Proposition~\ref{prop:MILP} holds due to the exact ReLU representation in \eqref{PWLPF1}--\eqref{PWLPF3} and the affine mappings in \eqref{PWLPF4}--\eqref{PWLPF6}. The latter follows from the transformation from $\bbz^2$ to $\bbz^3$ in our N-PWL model. The complex-valued equality in \eqref{PWLPF6} represents two real affine equalities, with admittance coefficients fixed per the discussions for \eqref{powerflow}.

Using the equivalent N-PWL PF representation in~\eqref{PWLPF}, the DNR and ROP problems in \eqref{DNR} and \eqref{ROP} can be approximated by MILP formulations. Note that the binary-continuous products in \eqref{actual_flow}, with $\epsilon_{ij}$ and bounded $\widehat S_{ij}$, are known \textit{equivalent to} the \emph{linear} McCormick constraints~\cite{mccormick1976computability}. In addition, 
the constraints on apparent power limits in \eqref{OPF4} are the \textit{sole nonlinear} inequalities in the two problems, making both mixed-integer quadratically constrained programs (MIQCPs).
It is very common to adopt a polyhedral approximation for this quadratic constraint in \eqref{OPF4}, as detailed in \cite{pareek2018piecewise}, where the resulting linear inequalities incur negligible approximation error of the line rating (below 0.625\%). These two transformations complete the MILP reformulations for both DNR and ROP, whose performance will be evaluated in the next section.

\section{Numerical Studies} \label{sec:sim}
We validate the proposed N-PWL PF model using the \textit{unbalanced} IEEE 123-bus multi-phase feeder~\cite{schneider2017analytic} for both PF modeling and topology optimization tasks. The N-PWL model training has been conducted in PyTorch with the Adam optimizer on a regular laptop with Intel\textsuperscript{\textregistered} CPU @ 2.70 GHz, 32 GB RAM, and NVIDIA\textsuperscript{\textregistered} RTX 3070 Ti GPU @ 8 GB VRAM. We formulated DNR and ROP problems based on the N-PWL PF and LinDistFlow (LDF)~\cite{baran1989optimal} models in Python. In addition, we used Gurobi~\cite{gurobi} to solve the resultant MILPs, while using BONMIN~\cite{bonami2008algorithmic} for their nonconvex AC counterparts in \eqref{DNR} and \eqref{ROP}. The network configuration decisions obtained from these various optimization methods are subsequently evaluated using a multi-phase PF model in OpenDSS~\cite{opendss_software}.
Throughout the simulations, bus 150r serves as the reference bus with $V_{\mathrm{ref}} = 1.05$ p.u. and the standard balanced three-phase angles.

For DNR/ROP formulations, we have set nine controllable branches, as summarized in Table~\ref{tab:SW}, which includes all eight switches (SW1--SW8) in the original test feeder. Note that the nominal statuses therein are used to determine the branch-status changes. For the selection of controllable branches, two original switches, namely SW1 and SW6, are set to \emph{Fixed Closed} to preserve network connectivity, while SW8 is set to \emph{Fixed Open} and excluded from optimization due to its unique three-to-single-phase connection. The remaining five original switches together with four new switches that are normally open three-phase ties (SW9--SW12) result in a total of nine controllable switches. Under a budget of up to six branch switchings, enforcing connectivity and radiality leads to 63 admissible network configurations, including the nominal one.

\begin{table}[t!]
\caption{\small Switch settings for the IEEE 123-Bus Feeder}
\centering
%\begin{center}
\begin{tabular}{c|c|c|c|c}
\Xhline{3\arrayrulewidth}
 & From--To & Phases & \makecell[c]{Nominal\\Status} & Control type \\ \hline
SW1   & 150r--149  & 3           & Closed & Fixed Closed \\
SW2   & 13--152  & 3           & Closed & Controllable  \\
SW3   & 18--135  & 3           & Closed & Controllable  \\
SW4   & 60--160  & 3           & Closed & Controllable  \\
SW5   & 97--197  & 3           & Closed & Controllable  \\
SW6   & 61--61s  & 3           & Closed & Fixed Closed \\
SW7   & 151--300  & 3           & Open   & Controllable  \\
SW8   & 54.1--94.1 & 1 & Open   & Fixed Open \\
SW9  & 54--93   & 3           & Open   & Controllable (Added)  \\
SW10 & 25--95   & 3           & Open   & Controllable (Added)  \\
SW11 & 35--64   & 3           & Open   & Controllable (Added)  \\
SW12 & 52--108  & 3           & Open   & Controllable (Added)  \\
\Xhline{3\arrayrulewidth}
\end{tabular}\label{tab:SW}
%\end{center}
\end{table}

To construct an N-PWL dataset covering a broad operating range, we generate 20,000 valid samples by applying 2,000 distinct load profiles to 10 representative radial topologies.
First, we obtain diverse load profiles from three loading cases: i) normal loading (40\%), ii) heavy loading (30\%), and iii) phase-imbalanced loading (30\%). These cases vary the loading magnitude and degree of phase imbalance, with the latter quantified by the following metric:
$$ \mathrm{PI}=\frac{\max_{\phi\in\{a,b,c\}}\Large||S^\phi|-|S^{\mathrm{avg}}|\Large|}{|S^{\mathrm{avg}}|}\times100\% $$
where $S^\phi$ and $S^{\mathrm{avg}}$ denote the phase-$\phi$ and average-phase loading, respectively.
The normal-loading case randomly perturbs the nodal active/reactive demands within [0.7,~1.3] of their nominal values with $\mathrm{PI}<15\%$, while the heavy-loading case uses a higher range of [1.2,~1.7] with the same $\mathrm{PI}$ bound. In contrast, the phase-imbalanced case retains the normal-loading perturbation range but uses a $\mathrm{PI}$ range of 10--40\% to capture asymmetries in PF outputs across phases.
Second, for topology coverage, the 10 representative topologies are selected such that each of the nine controllable branches in Table~\ref{tab:SW} is energized at least once.
This selection avoids enumerating all admissible configurations for training, leveraging the topology adaptivity established in Proposition~\ref{prop:topology}.
Finally, only converged ground-truth multi-phase PF solutions from OpenDSS satisfying the operating limits in \eqref{bound} are retained in the dataset. The corresponding target labels are $[\bbrho;\bbpi]$, potential branch flows, and nodal injections, all in multi-phase.

The backpropagation for N-PWL training uses $\lambda=10$ for the loss in \eqref{update} with a total of $20$e$3$ epochs and a learning rate of $2$e$-4$. This $\lambda$ value is selected through hyperparameter tuning to prioritize the prediction of PF outputs $[\bbz^{\mathrm{pf}};\bbz^{\mathrm{inj}}]$ while accurately approximating $[\bbrho;\bbpi]$. In addition, the 2,000 load profiles have a 90\%-10\% division for training and testing stages.
Unlike the conventional testing stage focused on unseen load profiles, we have tested the N-PWL model beyond the training topologies,  covering \textit{all 63 admissible network configurations} for our 123-bus test feeder.

\begin{figure}[t!]
	\centering
	\subfloat[Average error]{\includegraphics[scale=0.25]{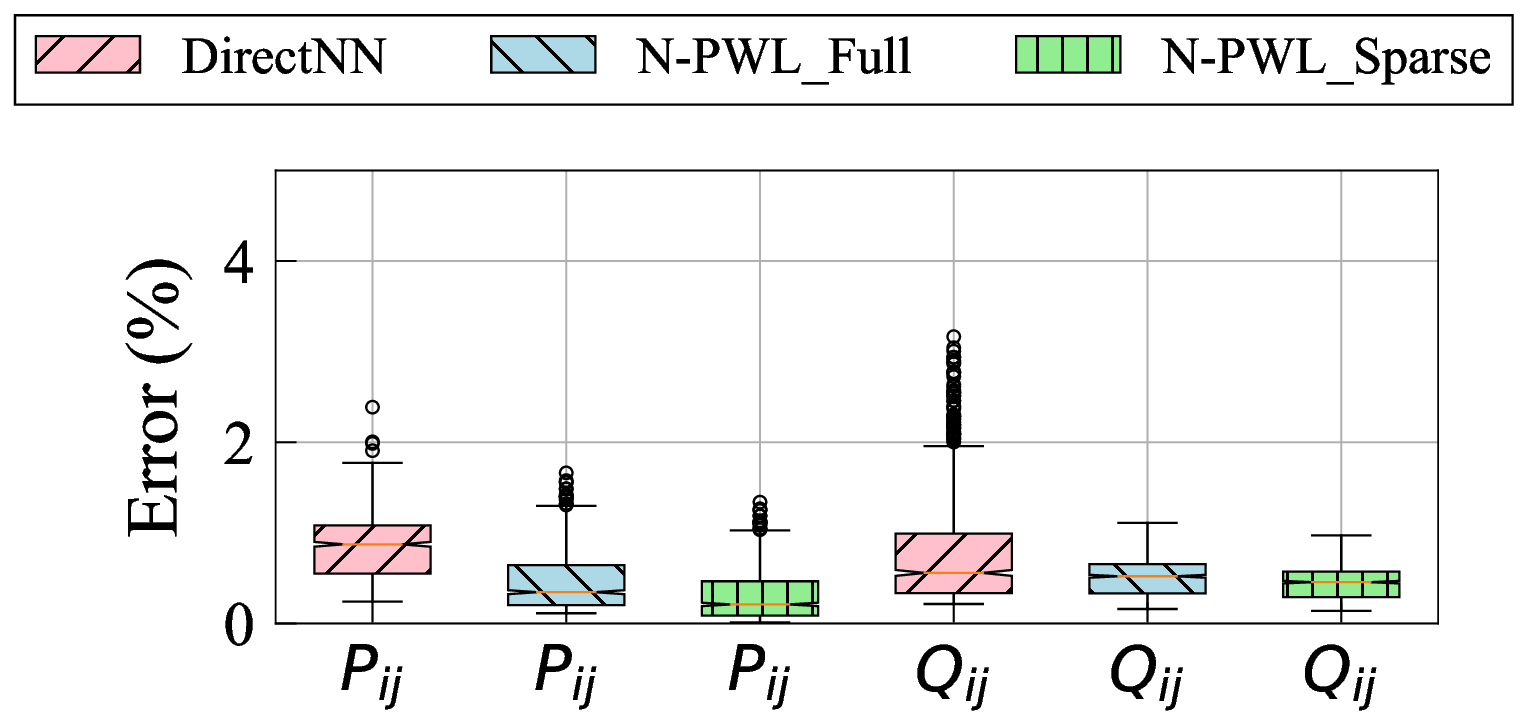}\label{avg_123}}
	\quad
	\subfloat[Maximum error]{\includegraphics[scale=0.25]{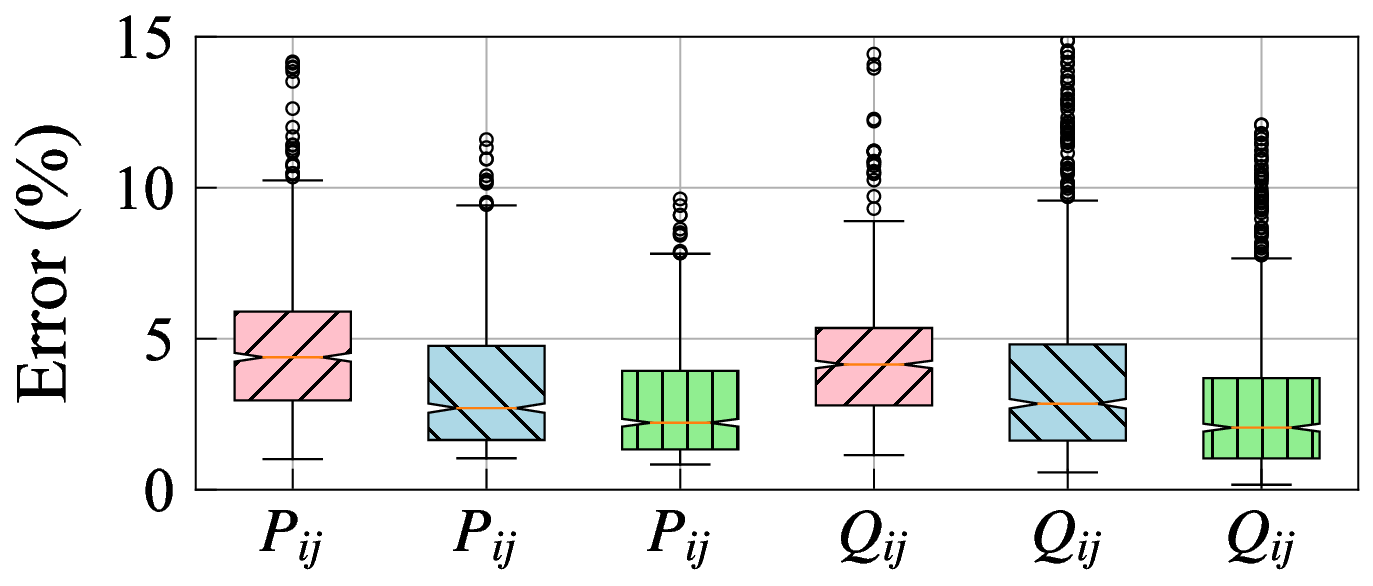}\label{max_123}}
    \quad
    \subfloat[Injection error]{\includegraphics[scale=0.25]{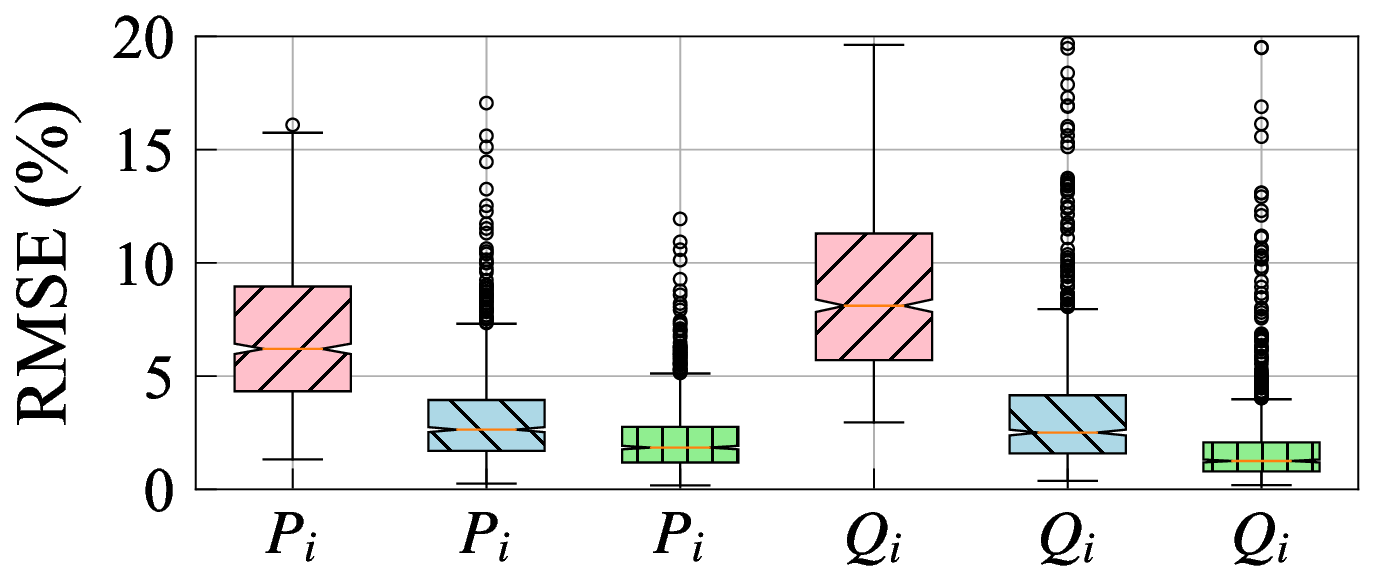}\label{inj_123}}
	\caption[]{\small Comparisons of the (a) average and (b) maximum error in approximating the phase-wise potential branch flows and (c) RMSE in predicting the phase-wise nodal injection vectors for DirectNN, N-PWL\_Full and N-PWL\_Sparse methods using the 123-bus feeder system.}\label{ACPF_123}
\end{figure}

\subsection{N-PWL Power Flow Modeling Performance}\label{sec:pf}

We compare our proposed N-PWL model against a baseline neural PF model. The baseline, termed as DirectNN, uses a two-layer NN to directly map multi-phase voltage state $\bbx=[\bbV;\bbtheta]$ to all PF outputs in $[\bbz^{\mathrm {pf}};\bbz^{\mathrm {inj}}]$. This architecture follows the general end-to-end PF modeling approaches in~\cite{kody2022modeling,chevalier2022global}. Note that this baseline \textit{cannot} support varying topologies, and thus one individual DirectNN model has to be trained for each of the 10 representative topologies. For our N-PWL model, we also consider two designs: N-PWL\_Full with fully connected trainable parameters, and N-PWL\_Sparse using the low-complexity design developed in Section~\ref{sec:sparse}. To ensure consistency in comparisons, all three models use $K=275$ hidden neurons, with N-PWL\_Sparse distributing them uniformly across five areas.

We first present the PF approximation errors over these 10 representative topologies used in training, with box plots shown in Fig.~\ref{ACPF_123}. Specifically, Figs.~\ref{ACPF_123}(a)-(b) compare the average and maximum error percentages in predicting $\bbz^{\mathrm {pf}}$ based on a normalization basis of branch ratings, while Fig.~\ref{ACPF_123}(c) shows the root mean square error (RMSE) in predicting $\bbz^{\mathrm {inj}}$. Each box plot indicates the median values as midlines along with the first (Q1) and third (Q3) quartiles as boxes, whereas the top bars isolate the outliers with values above 1.5 times the interquartile range (Q1-Q3) from the box.

Our two N-PWL models have clearly outperformed the DirectNN baseline in predicting both types of PF outputs. This improvement is especially significant for predicting $\bbz^{\rm inj}$ in Fig.~\ref{ACPF_123}(c).
While the N-PWL models preserve PF branch-node consistency, the DirectNN model neglects the coupling between $\bbz^{\rm pf}$ and $\bbz^{\rm inj}$, leading to larger injection errors.
Figs.~\ref{ACPF_123}(a)--(b) also show that our N-PWL models have better accuracy in approximating $\bbz^{\rm pf}$, thanks to their layer design that approximates the shared nonlinear terms $[\bbrho;\bbpi]$.
Between the two N-PWL models, N-PWL\_Sparse achieves slightly lower errors while retaining the linear parameter scaling established in Proposition~\ref{prop:sparse}.
In short, our proposed N-PWL models have improved PF modeling accuracy through PF-consistent neural designs, including a sparsified variant.

\begin{figure}[t!]
	\centering
	\subfloat[Average error]{\includegraphics[scale=0.25]{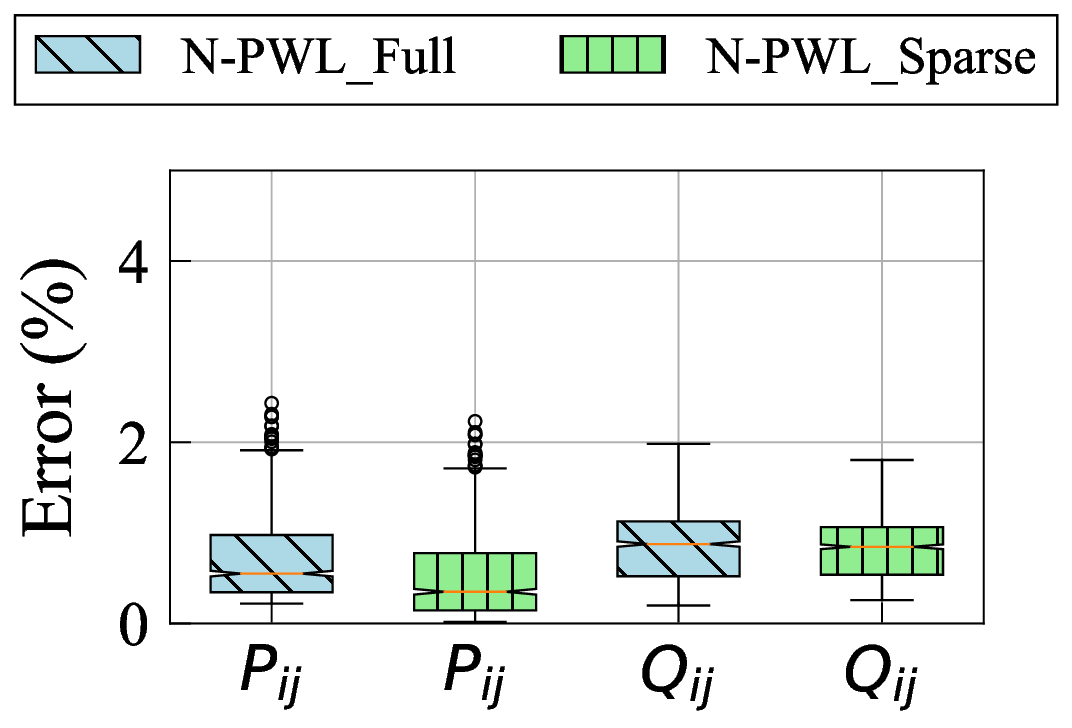}\label{avg_topo_123}}
	%\quad
	\subfloat[Maximum error]{\includegraphics[scale=0.25]{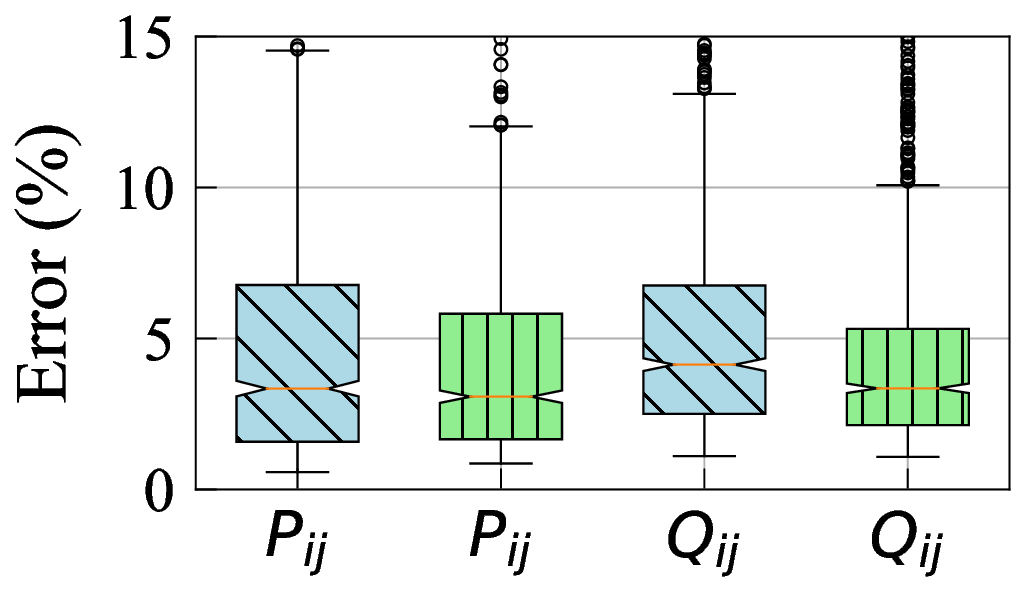}\label{max_topo_123}}
    \quad
    \subfloat[Injection error]{\includegraphics[scale=0.25]{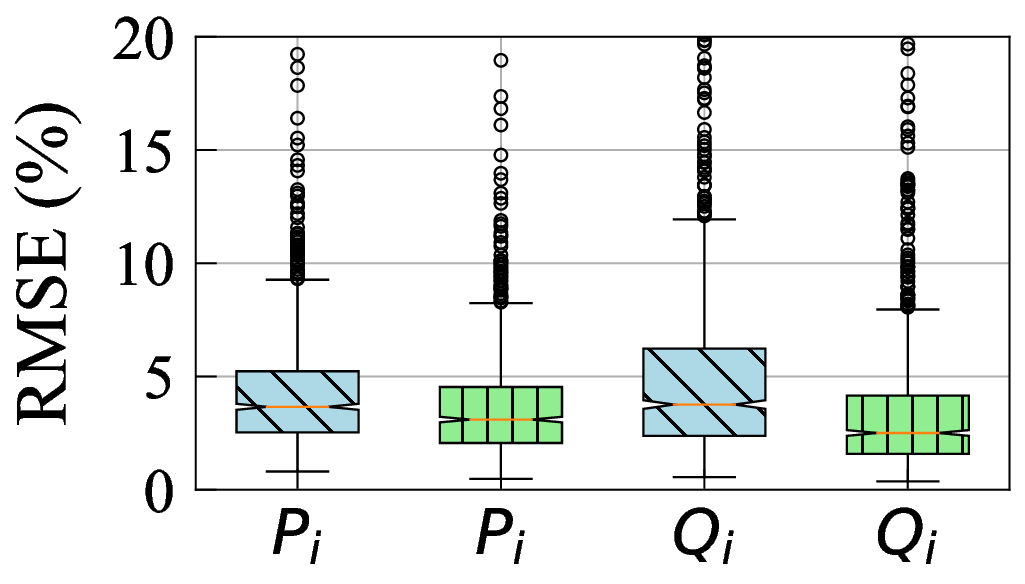}\label{inj_topo_123}}
	\caption[]{\small Comparisons of the (a) average and (b) maximum errors in approximating the phase-wise potential branch flows and (c) RMSE in predicting the phase-wise nodal injection vectors for N-PWL\_Full and N-PWL\_Sparse over unseen network configurations.}
	\label{ACPF_topology}
\end{figure}

We next evaluate the \textit{topology adaptivity} of the N-PWL models on the remaining 53 network configurations \textit{unseen} by the training stage, with the same error metrics as before.
As the DirectNN model requires topology-specific training, it is excluded from this comparison.
But our N-PWL models can represent any new topology just by updating $\bbW^\psi(\bbepsilon)$ in the final layer. Fig.~\ref{ACPF_topology}(a) has confirmed that both models maintain the same level of average errors even for unseen topologies. While Fig.~\ref{ACPF_topology}(b) shows a slight increase in maximum errors, this small degradation is expected for our setting of \textit{zero-shot} topology transfer.
Together, these results demonstrate that both N-PWL models are suited for subsequent topology-optimization studies, but only N-PWL\_Sparse will be used due to its high accuracy and low model complexity.

\subsection{N-PWL PF for Solving DNR Problem}\label{sec:simdnr}
We compare the N-PWL model selected in Section~\ref{sec:pf} for solving the DNR problem in \eqref{DNR} with LinDistFlow (LDF) and AC-PF models. Both N-PWL and LDF models yield MILP formulations for DNR, with the former using \eqref{PWLPF} and the latter the conventional linearized PF model in \cite{baran1989optimal}.
As for AC-DNR, it retains the nonlinear PF equations in \eqref{powerflow}, resulting in a nonconvex mixed-integer nonlinear program (MINLP).
%{\hao the next sentence doesn't flow well}
Solving this MINLP can be numerically challenging, as branch-and-bound over binary topology decisions involves repeated local NLP solves using IPOPT~\cite{wachter2006implementation}.
%Solving this problem combines NLP solves using the local interior-point optimizer (IPOPT)~\cite{wachter2006implementation} with branch-and-bound over binary topology decisions, potentially leading to numerical difficulties.
%
For all methods, we consider switching budgets $\alpha\in\{2,4,6\}$ and 80 normal-loading profiles from the testing dataset to assess average performance. As DNR seeks to identify the topology decisions, we extract the $\bbepsilon$ solution from each method and re-evaluate the resulting topology using \textit{multi-phase PF solutions} from OpenDSS. This AC re-evaluation provides a fair comparison of the \textit{realized performance} across DNR methods.

We evaluate the DNR methods based on three criteria: optimality, feasibility, and computation time.
For the optimality comparison, AC-DNR serves as the baseline with its realized power loss $P_{\mathrm{loss}}$ computed using \eqref{DNR_a}. The ratio $P_{\mathrm{loss}}/P_{\mathrm{loss}}^m$ reports the sub-optimality of each competing method $m$ relative to AC-DNR, with a higher value indicating better optimality (100\% for AC-DNR itself).
As for feasibility, we consider two metrics: i) \textit{solver-failure rate} of selected topologies for which OpenDSS fails to converge, and ii) \textit{constraint-violation rate} averaged over converged topologies.
Last, we report computation time as the optimization solve time for each DNR method, excluding model building and AC re-evaluation.

\begin{figure}[t!]
	%\begin{center}
	\centering
    \includegraphics[scale=0.33]{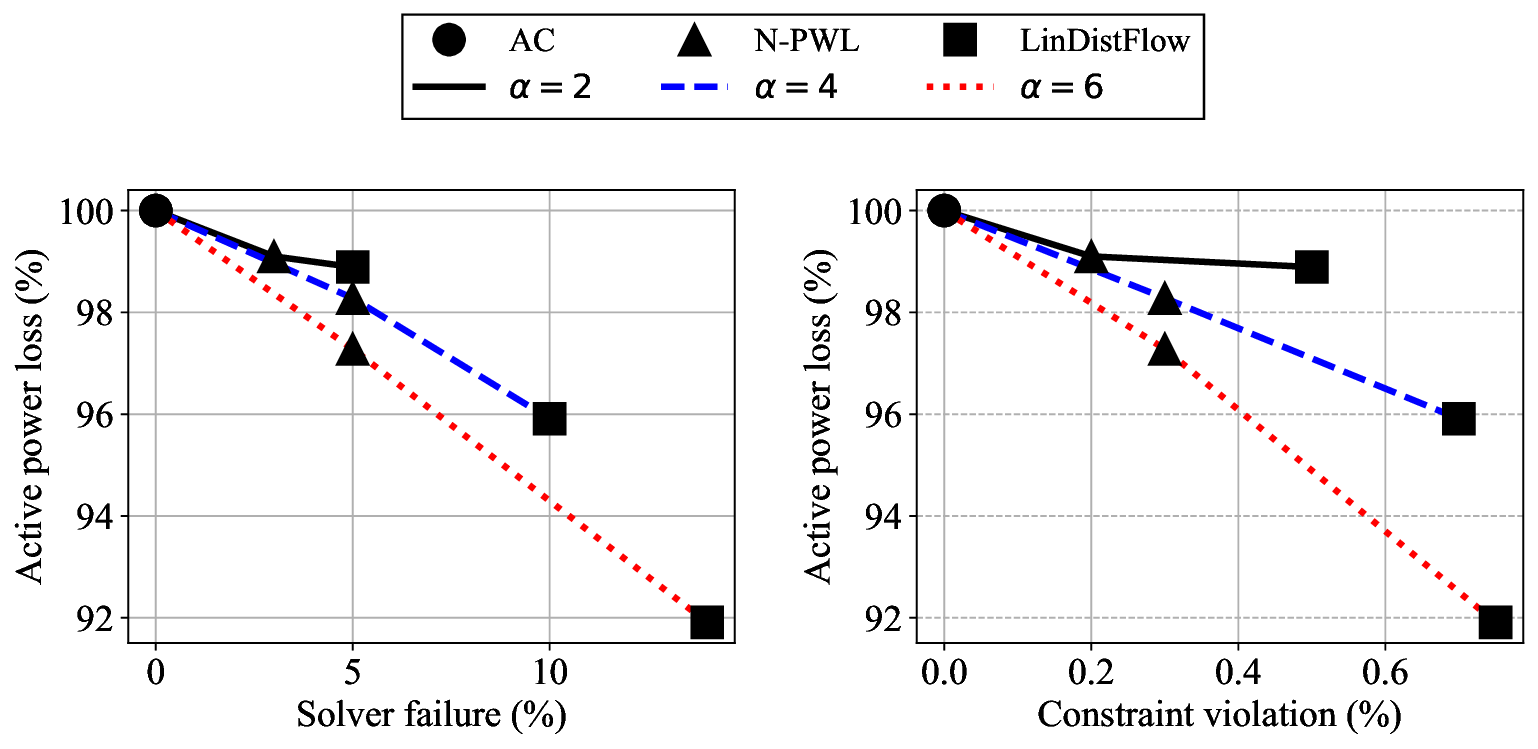}
	\caption{\small Comparisons of the DNR optimality versus feasibility performance using the 123-bus feeder system.} \label{DNR_sim}
	%\end{center}
\end{figure}

\begin{figure}[t!]
	%\begin{center}
	\includegraphics[scale=0.3]{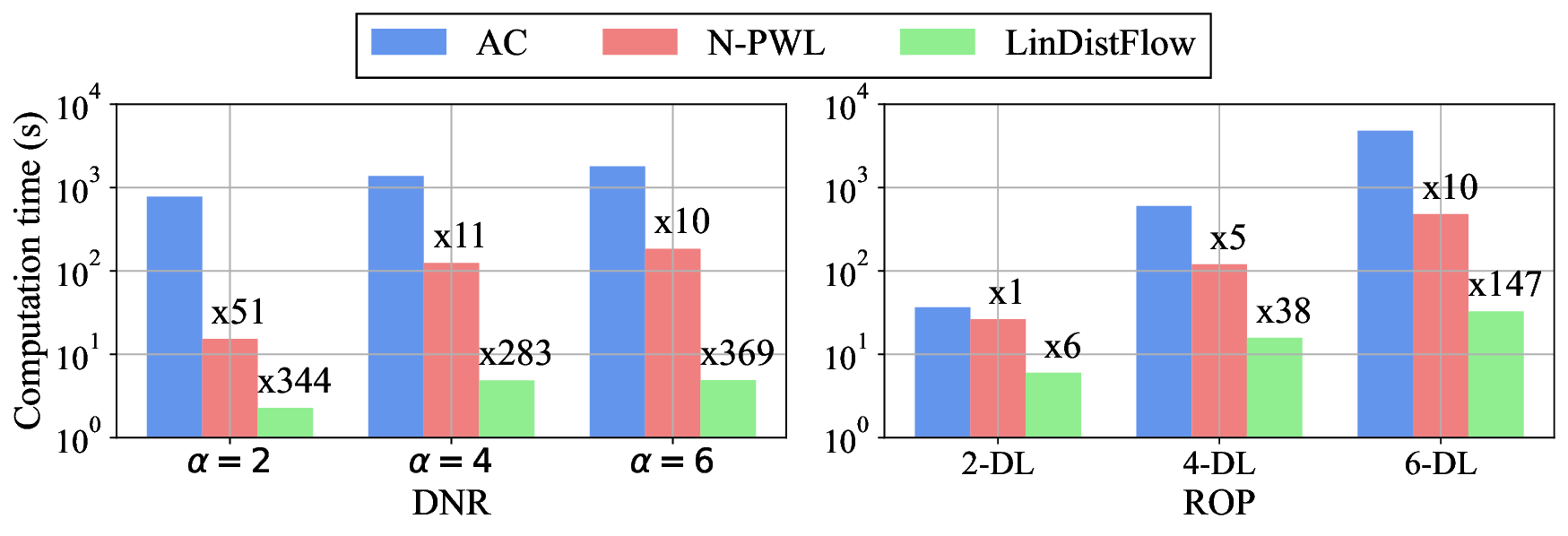}
	\caption{\small Comparisons of DNR and ROP computation times on a logarithmic scale using the IEEE 123-bus feeder, with speedup factors relative to their AC-PF-based counterparts.} %{\hao can we change the scaling of yaxis to make the diff between AC and N-PWL larger?}}
    \label{ComputationTime}
	%\end{center}
\end{figure}

Fig.~\ref{DNR_sim} plots the performance trade-offs between optimality and each of the two feasibility metrics, with better performance toward the upper-left corner.
As expected, AC-DNR achieves the best overall performance. But our proposed PWL-DNR approaches this baseline with an optimality ratio above 97\%, with a noticeable improvement over LDF-DNR. This comparison also holds for the two feasibility metrics, and it becomes even more pronounced as the switching budget $\alpha$ increases.
Moreover, the computation-time comparison in Fig.~\ref{ComputationTime} confirms that PWL-DNR is much more efficient than AC-DNR, requiring approximately one-tenth of its solve time at larger switching budgets.
While LDF-DNR is the fastest, its poorer AC feasibility limits its suitability for topology optimization.
Together, these results demonstrate that our proposed N-PWL model efficiently identifies high-quality topology decisions.

\subsection{Applications to the ROP}\label{sec:simrop}
We further compare the N-PWL model for solving the ROP in \eqref{ROP} with LDF and AC-PF models. The N-PWL and LDF models yield multi-period MILP formulations, while AC-ROP is formulated as a nonconvex MINLP. For solving the latter, we use the same MINLP solver settings as AC-DNR, but the time-coupling makes it more complicated to solve.
For a fair comparison, we also use an AC re-evaluation step to assess the restoration-order decisions $\bbepsilon_t$ produced by each of the three ROP methods. Specifically, we evaluate the \textit{realized performance} by re-solving AC-ROP with fixed $\bbepsilon_t$. This fixed-order ROP reduces the original MINLP to an NLP, in which only the continuous operating and load-service variables are optimized using IPOPT. Although IPOPT does not guarantee global optimality, the resulting AC-feasible solutions provide a baseline for performance comparisons.

We evaluate the ROP methods using the same procedures as in DNR for different numbers of damaged lines (DLs).
For the optimality comparison, AC-ROP serves as the basis with its realized total energy served $E_{\mathrm{srv}}$ computed using \eqref{ROP_a}. The ratio $E_{\mathrm{srv}}^m/E_{\mathrm{srv}}$ reports the sub-optimality of each competing method $m$ relative to AC-ROP, with a higher value indicating better optimality (100\% for AC-ROP itself). The metrics on feasibility and computation time also follow those used for DNR.
To examine the impact of restoration complexity, we consider 2-, 4-, and 6-DL cases on the nominal network configuration. With a restoration budget of $\eta=1$ per period, the corresponding restoration horizons are $T\in\{2,4,6\}$. We use the same 80 normal-loading profiles as in the DNR study to assess average performance.

\begin{figure}[t!]
	%\begin{center}
	\includegraphics[scale=0.33]{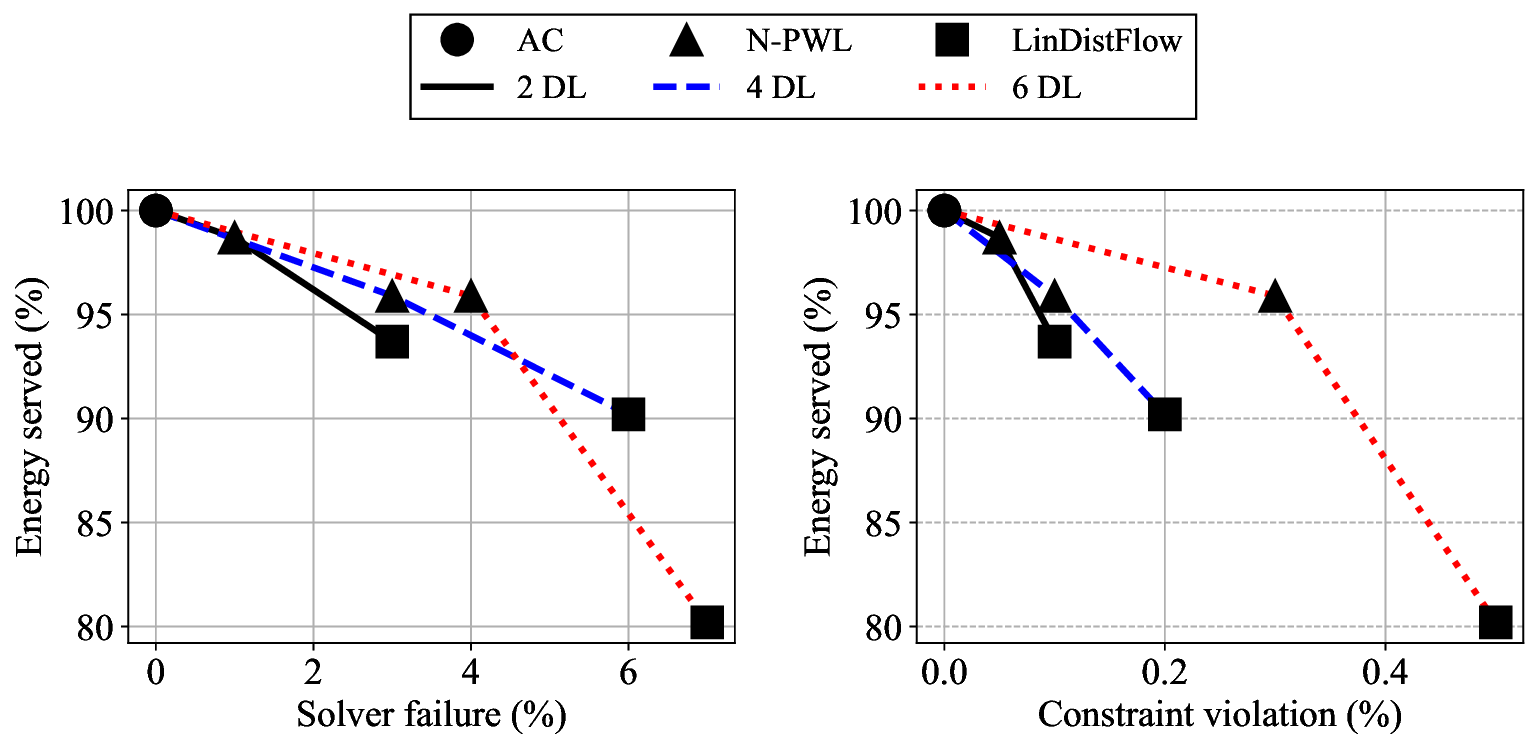}
	\caption{\small Comparisons of the ROP optimality versus feasibility performance using the 123-bus feeder system.} \label{ROP_sim}
	%\end{center}
\end{figure}

The computation-time comparison in Fig.~\ref{ComputationTime} shows the computational advantage of PWL-ROP as restoration complexity grows. For the 2-DL case, the small number of restoration decisions allows even AC-ROP to be solved faster than single-period AC-DNR. However, the 6-DL case involves a substantially larger search space, for which AC-ROP takes approximately three times as long as AC-DNR. Yet, PWL-ROP provides an order-of-magnitude reduction in computation time relative to AC-ROP, demonstrating its better scalability for larger DL cases.
Moreover, PWL-ROP exhibits favorable optimality and feasibility, consistent with the DNR results. Fig.~\ref{ROP_sim} shows that it approaches the AC-ROP baseline with an optimality ratio above 96\% and outperforms LDF-ROP in both optimality and feasibility, particularly for larger DL cases.
Together, these results demonstrate that our proposed N-PWL model enables efficient computation of high-quality restoration-order decisions, providing effective binary search for more complex, larger-DL cases.

\section{Conclusions} \label{sec:con}
We developed a topology-adaptive, low-complexity neural PWL (N-PWL) power-flow (PF) model for distribution grid topology optimization. Our sequential layer design directly incorporates branch status variables into the PF mapping from branch flows to nodal injections. This design not only preserves PF branch-node consistency but also facilitates the representation of unseen topologies without retraining. We further reduced the N-PWL model complexity through area-based sparsification of its trainable layers, achieving linear scaling of NN parameters with feeder size. The resulting scalable N-PWL model was utilized for mixed-integer linear programming (MILP) reformulations of feeder topology optimization problems including DNR and ROP. Numerical studies on the unbalanced, multi-phase IEEE 123-bus feeder demonstrated improved PF prediction accuracy over existing end-to-end baselines and, more importantly, the accurate representation of unseen topologies. For DNR and ROP, our model improved AC feasibility over linearized PF models and reduced computation time compared with AC-PF based nonlinear formulations, particularly under a large number of topology decisions.

%We proposed a novel neural PWL (N-PWL) model that accurately approximates nonlinear AC-PF equations while explicitly preserving the underlying branch-node coupling. We put forth a low-complexity design using the area-based sparsification to attain linear parameter scaling for large systems. By directly embedding the binary topology variables into the layers, our N-PWL can adapt to varying topologies unseen by the training stage, while maintaining the same PF accuracy. The resultant PWL-PF approximation has been successfully used for the MILP reformulations of grid topology optimization tasks such as distribution network reconfiguration (DNR) and restoration ordering problems (ROP). Numerical tests on the IEEE 123-bus feeder corroborate its excellent modeling accuracy, as well as the scalability and computational efficiency for topology optimization, particularly for a larger number of binary decisions. 

Future directions include investigating GPU-based solvers for MILP acceleration and extending our N-PWL modeling framework to multi-energy systems and related applications, such as power system decision-making coupled with water networks and other natural resources. In addition, we could explore broader problems in distribution grid monitoring and operations to address its limited sensing and control capabilities.

\begin{comment}
Exciting future directions open up for our GenNN-based linearizable PF models, such as extensions to modeling distribution power flow or carbon emission flow. We can also consider general grid monitoring and operation tasks by accounting for limited sensing and control capabilities. Moreover, implementing the PWL-based grid optimization solutions could benefit from reducing the complexity of the MILP reformulation.
\end{comment}

\ifCLASSOPTIONcaptionsoff
\newpage
\fi
	
\bibliographystyle{IEEEtran}
\bibliography{Ref.bib}

@inproceedings{van2011vehicle,
  title={Vehicle routing for the last mile of power system restoration},
  author={Van Hentenryck, Pascal and Coffrin, Carleton and Bent, Russell and others},
  booktitle={2011 Power Systems Computation Conference (PSCC)},
  year={2011}
}

@inproceedings{kody2022modeling,
  author = {Kody, Alyssa and Chevalier, Samuel and Chatzivasileiadis, Spyros and Molzahn, Daniel},
  title = {Modeling the {AC} power flow equations with optimally compact neural networks: Application to unit commitment},
  booktitle = {2022 Power Systems Computation Conference (PSCC)},
  year = {2022}
}

@misc{gurobi,
  author = {{Gurobi Optimization, LLC}},
  title = {{Gurobi Optimizer Reference Manual}},
  year = 2023,
  url = "https://www.gurobi.com"
}

@article{grimstad2019relu,
  title={Re{LU} networks as surrogate models in mixed-integer linear programs},
  author={Grimstad, Bjarne and Andersson, Henrik},
  journal={Computers \& Chemical Engineering},
  volume={131},
  pages={106580},
  year={2019},
  publisher={Elsevier}
}

@article{trodden2013optimization,
  title={Optimization-based islanding of power networks using piecewise linear {AC} power flow},
  author={Trodden, Paul A and Bukhsh, Waqquas Ahmed and Grothey, Andreas and McKinnon, Ken IM},
  journal={IEEE Trans. Power Systems},
  volume={29},
  number={3},
  pages={1212--1220},
  year={2014},
  publisher={IEEE}
}

@book{griva2009linear,
  title={Linear and nonlinear optimization},
  author={Griva, Igor and Nash, Stephen G and Sofer, Ariela},
  volume={108},
  year={2009},
  publisher={Siam}
}

@article{mccormick1976computability,
  title={Computability of global solutions to factorable nonconvex programs: {P}art {I}—{C}onvex underestimating problems},
  author={McCormick, Garth P},
  journal={Mathematical programming},
  volume={10},
  number={1},
  pages={147--175},
  year={1976},
  publisher={Springer}
}

@article{liu2023topology,
  title={{Topology-aware graph neural networks for learning feasible and adaptive AC-OPF solutions}},
  author={Liu, Shaohui and Wu, Chengyang and Zhu, Hao},
  journal={IEEE Trans. Power Systems},
  volume={38},
  number={6},
  pages={5660--5670},
  year={2023},
  publisher={IEEE}
}

@article{chen2021data,
  title={Data-driven piecewise linearization for distribution three-phase stochastic power flow},
  author={Chen, Jiaqi and Wu, Wenchuan and Roald, Line A},
  journal={IEEE Trans. Smart Grid},
  volume={13},
  number={2},
  pages={1035--1048},
  year={2022},
  publisher={IEEE}
}

@article{simpson2017theory,
  title={A theory of solvability for lossless power flow equations—{P}art {I}: {F}ixed-point power flow},
  author={Simpson-Porco, John W},
  journal={IEEE Trans. Control of Network Systems},
  volume={5},
  number={3},
  pages={1361--1372},
  year={2018},
  publisher={IEEE}
}

@article{sanchez2014hierarchical,
  title={Hierarchical spectral clustering of power grids},
  author={S{\'a}nchez-Garc{\'\i}a, Rub{\'e}n J and Fennelly, Max and Norris, Se{\'a}n and Wright, Nick and Niblo, Graham and Brodzki, Jacek and Bialek, Janusz W},
  journal={IEEE Trans. Power Systems},
  volume={29},
  number={5},
  pages={2229--2237},
  year={2014},
  publisher={IEEE}
}

@article{anderson2020strong,
  title={Strong mixed-integer programming formulations for trained neural networks},
  author={Anderson, Ross and Huchette, Joey and Ma, Will and Tjandraatmadja, Christian and Vielma, Juan Pablo},
  journal={Mathematical Programming},
  volume={183},
  number={1},
  pages={3--39},
  year={2020},
  publisher={Springer}
}

@inproceedings{cho2023topology,
  title={Topology-Aware Piecewise Linearization of the {AC} Power Flow Through Generative Modeling},
  author={Cho, Young-ho and Zhu, Hao},
  booktitle={2023 North American Power Symposium (NAPS)},
  pages={1--6},
  year={2023},
  organization={IEEE}
}

@article{trunk1979problem,
  title={A problem of dimensionality: A simple example},
  author={Trunk, Gerard V},
  journal={IEEE Trans. Pattern Analysis and Machine Intelligence},
  number={3},
  pages={306--307},
  year={1979},
  publisher={IEEE}
}

@article{jalving2024physics,
  title={Physics-informed machine learning with optimization-based guarantees: Applications to {AC} power flow},
  author={Jalving, Jordan and Eydenberg, Michael and Blakely, Logan and Castillo, Anya and Kilwein, Zachary and Skolfield, J Kyle and Boukouvala, Fani and Laird, Carl},
  journal={International Journal of Electrical Power \& Energy Systems},
  volume={157},
  pages={109741},
  year={2024},
  publisher={Elsevier}
}

@article{liu2018data,
  title={Data-driven power flow linearization: A regression approach},
  author={Liu, Yuxiao and Zhang, Ning and Wang, Yi and Yang, Jingwei and Kang, Chongqing},
  journal={IEEE Trans. Smart Grid},
  volume={10},
  number={3},
  pages={2569--2580},
  year={2019},
  publisher={IEEE}
}

@article{ng2001spectral,
  title={On spectral clustering: Analysis and an algorithm},
  author={Ng, Andrew and Jordan, Michael and Weiss, Yair},
  journal={Advances in Neural Information Processing Systems},
  volume={14},
  year={2001}
}

@article{fattahi2018bound,
  title={A bound strengthening method for optimal transmission switching in power systems},
  author={Fattahi, Salar and Lavaei, Javad and Atamt{\"u}rk, Alper},
  journal={IEEE Trans. Power Systems},
  volume={34},
  number={1},
  pages={280--291},
  year={2019},
  publisher={IEEE}
}

@article{lin2024powerflownet,
  title={PowerFlowNet: Power flow approximation using message passing Graph Neural Networks},
  author={Lin, Nan and Orfanoudakis, Stavros and Cardenas, Nathan Ordonez and Giraldo, Juan S and Vergara, Pedro P},
  journal={International Journal of Electrical Power \& Energy Systems},
  volume={160},
  pages={110112},
  year={2024},
  publisher={Elsevier}
}

@article{pan2020deepopf,
  title={Deepopf: A deep neural network approach for security-constrained {DC} optimal power flow},
  author={Pan, Xiang and Zhao, Tianyu and Chen, Minghua and Zhang, Shengyu},
  journal={IEEE Trans. Power Systems},
  volume={36},
  number={3},
  pages={1725--1735},
  year={2020},
  publisher={IEEE}
}

@article{chen2023end,
  title={End-to-end feasible optimization proxies for large-scale economic dispatch},
  author={Chen, Wenbo and Tanneau, Mathieu and Van Hentenryck, Pascal},
  journal={IEEE Trans. Power Systems},
  year={2023},
  publisher={IEEE}
}

@article{chevalier2022global,
  title={{Global performance guarantees for neural network models of AC power flow}},
  author={Chevalier, Samuel and Chatzivasileiadis, Spyros},
  journal={arXiv preprint arXiv:2211.07125},
  year={2022}
}

@article{buason2025sample,
  title={Sample-Based Piecewise Linear Power Flow Approximations Using Second-Order Sensitivities},
  author={Buason, Paprapee and Misra, Sidhant and Molzahn, Daniel K},
  journal={arXiv preprint arXiv:2501.13825},
  year={2025}
}

@article{jabr2012minimum,
  title={Minimum loss network reconfiguration using mixed-integer convex programming},
  author={Jabr, Rabih A and Singh, Ravindra and Pal, Bikash C},
  journal={IEEE Trans. Power systems},
  volume={27},
  number={2},
  pages={1106--1115},
  year={2012},
  publisher={IEEE}
}

@misc{low2025power,
  author       = {Steven H. Low},
  title        = {Power System Analysis: Analytical Tools and Structural Properties},
  year         = {2025},
  month        = apr,
  url          = {https://netlab.caltech.edu/assets/book/PSA/Low-PSA-v20250407.pdf}
}

@article{bazrafshan2017comprehensive,
  title={Comprehensive modeling of three-phase distribution systems via the bus admittance matrix},
  author={Bazrafshan, Mohammadhafez and Gatsis, Nikolaos},
  journal={IEEE Trans. Power Systems},
  volume={33},
  number={2},
  pages={2015--2029},
  year={2018},
  publisher={IEEE}
}

@inproceedings{gan2014convex,
  title={Convex relaxations and linear approximation for optimal power flow in multiphase radial networks},
  author={Gan, Lingwen and Low, Steven H},
  booktitle={2014 Power Systems Computation Conference (PSCC)},
  pages={1--9},
  year={2014},
  organization={IEEE}
}

@article{bernstein2018load,
  title={Load flow in multiphase distribution networks: Existence, uniqueness, non-singularity and linear models},
  author={Bernstein, Andrey and Wang, Cong and Dall’Anese, Emiliano and Le Boudec, Jean-Yves and Zhao, Changhong},
  journal={IEEE Trans. Power Systems},
  volume={33},
  number={6},
  pages={5832--5843},
  year={2018},
  publisher={IEEE}
}

@article{bahrami2024dynamic,
  title={Dynamic distribution network reconfiguration with generation and load uncertainty},
  author={Bahrami, Shahab and Chen, Yu Christine and Wong, Vincent WS},
  journal={IEEE Trans. Smart Grid},
  volume={15},
  number={6},
  pages={5472--5484},
  year={2024},
  publisher={IEEE}
}

@article{oh2025sequential,
  title={Sequential control of individual switches for real-time distribution network reconfiguration using deep reinforcement learning},
  author={Oh, Jae-Young and Oh, Seokhwa and Lee, Gyu-Sub and Yoon, Yong Tae and Jo, Seungchan},
  journal={IEEE Trans. Smart Grid},
  volume={16},
  number={5},
  pages={3666--3683},
  year={2025},
  publisher={IEEE}
}

@article{arif2017power,
  title={Power distribution system outage management with co-optimization of repairs, reconfiguration, and {DG} dispatch},
  author={Arif, Anmar and Wang, Zhaoyu and Wang, Jianhui and Chen, Chen},
  journal={IEEE Trans. Smart Grid},
  volume={9},
  number={5},
  pages={4109--4118},
  year={2018},
  publisher={IEEE}
}

@article{xu2024adaptive,
  title={Adaptive robust load restoration via coordinating distribution network reconfiguration and mobile energy storage},
  author={Xu, Ruipeng and Zhang, Cuo and Zhang, Daming and Dong, Zhao Yang and Yip, Christine},
  journal={IEEE Trans. Smart Grid},
  volume={15},
  number={6},
  pages={5485--5499},
  year={2024},
  publisher={IEEE}
}

@article{liu2021data,
  title={Data-driven-aided linear three-phase power flow model for distribution power systems},
  author={Liu, Yitong and Li, Zhengshuo and Zhou, Yu},
  journal={IEEE Trans. Power Systems},
  volume={37},
  number={4},
  pages={2783--2795},
  year={2022},
  publisher={IEEE}
}

@article{li2022data,
  title={A data-driven linear optimal power flow model for distribution networks},
  author={Li, Penghua and Wu, Wenchuan and Wang, Xiaoming and Xu, Bin},
  journal={IEEE Trans. Power Systems},
  volume={38},
  number={1},
  pages={956--959},
  year={2023},
  publisher={IEEE}
}

@inproceedings{chen2023topology,
  title={Topology-adaptive piecewise linearization for three-phase power flow calculations in distribution grids},
  author={Chen, Jiaqi and Roald, Line A},
  booktitle={2023 North American Power Symposium (NAPS)},
  pages={1--6},
  year={2023},
  organization={IEEE}
}

@article{schneider2017analytic,
  title={Analytic considerations and design basis for the {IEEE} distribution test feeders},
  author={Schneider, Kevin P and Mather, BA and Pal, Bikash Chandra and Ten, C-W and Shirek, Greg J and Zhu, Hao and Fuller, Jason C and Pereira, Jos{\'e} Luiz Rezende and Ochoa, Luis F and de Araujo, Leandro Ramos and others},
  journal={IEEE Trans. Power Systems},
  volume={33},
  number={3},
  pages={3181--3188},
  year={2018},
  publisher={IEEE}
}

@article{baran1989network,
  title={Network reconfiguration in distribution systems for loss reduction and load balancing},
  author={Baran, Mesut and Wu, Felix F},
  journal={IEEE Trans. Power Delivery},
  volume={4},
  number={2},
  pages={1401--1407},
  year={1989},
  publisher={IEEE}
}

@article{pareek2018piecewise,
  title={Piecewise Linearization of Quadratic Branch Flow Limits by Irregular Polygon},
  author={Pareek, Parikshit and Verma, Ashu},
  journal={IEEE Trans. Power Systems},
  volume={33},
  number={6},
  pages={7301-7304},
  year={2018},
  publisher={IEEE}
}

@article{robbins2015optimal,
  title={Optimal tap setting of voltage regulation transformers in unbalanced distribution systems},
  author={Robbins, Brett A and Zhu, Hao and Dom{\'\i}nguez-Garc{\'\i}a, Alejandro D},
  journal={IEEE Trans. Power Systems},
  volume={31},
  number={1},
  pages={256--267},
  year={2016},
  publisher={IEEE}
}

@article{bazrafshan2019optimal,
  title={Optimal Power Flow With Step-Voltage Regulators in Multi-Phase Distribution Networks},
  author={Bazrafshan, Mohammadhafez and Gatsis, Nikolaos and Zhu, Hao},
  journal={IEEE Trans. Power Systems},
  volume={34},
  number={6},
  pages={4228-4239},
  year={2019},
  publisher={IEEE}
}

@article{bonami2008algorithmic,
  title={An algorithmic framework for convex mixed integer nonlinear programs},
  author={Bonami, Pierre and Biegler, Lorenz T and Conn, Andrew R and Cornu{\'e}jols, G{\'e}rard and Grossmann, Ignacio E and Laird, Carl D and Lee, Jon and Lodi, Andrea and Margot, Fran{\c{c}}ois and Sawaya, Nicolas and others},
  journal={Discrete optimization},
  volume={5},
  number={2},
  pages={186--204},
  year={2008},
  publisher={Elsevier}
}

@misc{opendss_software,
  title        = {{OpenDSS}: Open Source Distribution System Simulator},
  author       = {{Electric Power Research Institute (EPRI)}},
  year         = {2022},
  howpublished = {\url{https://epri.com}},
  note         = {Software available on SourceForge and GitHub}
}

@article{luan2024improved,
  title={Improved topology identification for distribution network with relatively balanced power supplies},
  author={Luan, Wenpeng and Xu, Da and Liu, Bo and Jiang, Wenqian and Feng, Li and Liu, Wenbin},
  journal={IET Energy Systems Integration},
  volume={6},
  number={2},
  pages={162--173},
  year={2024},
  publisher={Wiley Online Library}
}

@inproceedings{he2016deep,
  title={Deep residual learning for image recognition},
  author={He, Kaiming and Zhang, Xiangyu and Ren, Shaoqing and Sun, Jian},
  booktitle={Proceedings of the IEEE Conference on Computer Vision and Pattern Recognition},
  pages={770--778},
  year={2016}
}

@article{baran1989optimal,
  title={Optimal sizing of capacitors placed on a radial distribution system},
  author={Baran, Mesut and Wu, Felix F},
  journal={IEEE Trans. Power Delivery},
  volume={4},
  number={1},
  pages={735--743},
  year={1989},
  publisher={IEEE}
}

@article{wachter2006implementation,
  title={On the implementation of an interior-point filter line-search algorithm for large-scale nonlinear programming},
  author={W{\"a}chter, Andreas and Biegler, Lorenz T},
  journal={Mathematical programming},
  volume={106},
  number={1},
  pages={25--57},
  year={2006},
  publisher={Springer}
}

@article{duan2026two,
  title={A Two-Stage Topology-Adaptive Physics-Guided Power Flow Learning Approach for Power Distribution Networks},
  author={Duan, Mengzhao and Hu, Zhengyang and Chen, Junyu and Ding, Yibo and Zhao, Yuhong and Xu, Zhao},
  journal={IEEE Trans. Smart Grid},
  year={2026},
  publisher={IEEE}
}

@article{wang2024piecewise,
  title={{Piecewise affine power flow model for distribution network optimization using CIM aided data-driven approach}},
  author={Wang, Xiao and Zhang, Jing and Xu, Weimao and Tang, Aihong and Gao, Wenzhong},
  journal={Electric Power Systems Research},
  volume={237},
  pages={111026},
  year={2024},
  publisher={Elsevier}
}

@article{jia2024two,
  title={A two-stage approach for topology change-aware data-driven {OPF}},
  author={Jia, Yixiong and Wu, Xian and Yang, Zhifang and Wang, Yi},
  journal={IEEE Trans. Power Systems},
  volume={40},
  number={2},
  pages={1854--1867},
  year={2024},
  publisher={IEEE}
}

@article{zhou2021three,
  title={Three-phase unbalanced distribution network dynamic reconfiguration: A distributionally robust approach},
  author={Zhou, Anping and Zhai, Hefeng and Yang, Ming and Lin, You},
  journal={IEEE Trans. Smart Grid},
  volume={13},
  number={3},
  pages={2063--2074},
  year={2022},
  publisher={IEEE}
}

@inproceedings{singh2022joint,
  title={Joint grid topology reconfiguration and design of {Watt-VAR curves for DERs}},
  author={Singh, Manish K and Taheri, Sina and Kekatos, Vassilis and Schneider, Kevin P and Liu, Chen-Ching},
  booktitle={Proc.~IEEE Power \& Energy Society General Meeting (PESGM)},
  year={2022}
}

@article{taheri2026optimizing,
  title={Optimizing Parameters of the {LinDistFlow} Power Flow Approximation for Distribution Systems},
  author={Taheri, Babak and Gupta, Rahul K. and Molzahn, Daniel K.},
  journal={IEEE Trans. Smart Grid},
  volume={17},
  number={2},
  pages={1029--1042},
  year={2026},
  publisher={IEEE}
}

@article{qiu2023power,
  title={Power system restoration with renewable participation},
  author={Qiu, Feng and Zhang, Yichen and Yao, Rui and Du, Pengwei},
  journal={IEEE Trans. Sustainable Energy},
  volume={14},
  number={2},
  pages={1112--1121},
  year={2023},
  publisher={IEEE}
}

@article{wang2020radiality,
  title={On the radiality constraints for distribution system restoration and reconfiguration problems},
  author={Wang, Ying and Xu, Yin and Li, Jiaxu and He, Jinghan and Wang, Xiaojun},
  journal={IEEE Trans. Power Systems},
  volume={35},
  number={4},
  pages={3294--3296},
  year={2020},
  publisher={IEEE}
}

@inproceedings{chehade2026neo,
  title={{NEO-Grid: A Neural Approximation Framework for Optimization and Control in Distribution Grids}},
  author={Chehade, Mohamad and Zhu, Hao},
  booktitle={Proceedings of the 59th HICSS},
  year={2026}
}

\end{document}